\documentclass[12pt,a4paper]{article}
\usepackage[top=30mm, bottom=30mm, left=25mm, right=25mm]{geometry}

\usepackage[utf8]{inputenc}
\usepackage[T1]{fontenc}
\usepackage{lmodern}
\usepackage{amsmath,amssymb,amsfonts,amsthm}
\usepackage{mathtools}
\usepackage{booktabs}
\usepackage{graphicx}
\graphicspath{{figures/}{./}}
\usepackage{url}
\usepackage{microtype}
\usepackage{xcolor}
\usepackage{subcaption}
\usepackage{float}
\usepackage{placeins}
\newfloat{algorithm}{htbp}{loa}
\floatname{algorithm}{Algorithm}
\usepackage[authoryear,round]{natbib}
\usepackage[colorlinks=true, citecolor=blue, linkcolor=blue, urlcolor=blue]{hyperref}
\hypersetup{%
  pdftitle={Hill numbers for organisms and observers in space: Toward a democratic ecology},%
  pdfauthor={Shinto Eguchi, Osamu Komori, Yusuke Saigusa}%
}

\usepackage{bm}
\usepackage{tikz}
\usepackage{setspace}
\usetikzlibrary{arrows.meta,positioning}
\usepackage{tabularx}

\newtheorem{theorem}{Theorem}[section]
\newtheorem{proposition}[theorem]{Proposition}

\theoremstyle{remark}
\newtheorem*{remark}{Remark}

\newtheorem{lemma}[theorem]{Lemma}

\title{Hill numbers for organisms and observers in space:\\[2mm]
Toward a democratic ecology}

\author{Shinto Eguchi$^{1}$, Osamu Komori$^{2}$, and Yusuke Saigusa$^{3}$\\[8pt]
\small $^{1}$The Institute of Statistical Mathematics, Japan\\
\small $^{2}$Seikei University, Japan\\
\small $^{3}$Yokohama City University, Japan}
\date{}

\begin{document}
\maketitle

\vspace{5mm}

\begin{abstract}
Madhav Gadgil connected population dispersal and spatial crowding with
participatory ecological observation. We revisit these themes using Hill
numbers relative to resource and geographic reference measures. Under
logistic-type regulation and population-conserving movement, the total
population satisfies an exact Hill-number balance; a new space--time
identity characterizes mean abundance for periodic, temporally varying
carrying capacities. A fitness-driven dispersal model connects this
demographic Hill order to the exponent of nonlinear diffusion and
admits a Hill-number free-energy dissipation identity. The observational
counterpart separates biological distributions, recording events, and
species reports. We characterize the effect of shared observation effort
on regional diversity and establish an exact geographic transport limit
imposed by unvisited cells. Reproducible simulations and a descriptive
Watarase eBird checklist-location illustration demonstrate the distinction
between geographic concentration and ecological representation. The
results place spatial demographic imbalance and citizen-science recording
within a common reference-based language without conflating their
identification requirements.
\end{abstract}

\vspace{20mm}


\newpage
\section{Introduction}
\label{sec:introduction}

In his early study of dispersal, \citet{gadgil1971dispersal}
considered populations moving among sites with different
carrying capacities. His central observation was that dispersal
can leave some places overcrowded and others underoccupied,
reducing the population supported by a region as a whole.
Sites that are isolated or difficult to reach may consequently
remain less crowded. In his later work on the People's
Biodiversity Registers (PBRs), Gadgil and colleagues argued
that local people should participate in recording and interpreting
ecological knowledge \citep{gadgil2000participatory}.
We use the phrase \emph{democratic ecology} to describe the aspiration
that citizens participate in producing and interpreting ecological
knowledge, as exemplified by the PBRs.
Broad participation, however, does not itself guarantee representative
sampling: decisions about where and what to record may remain
spatially concentrated. The two projects invite us to compare the
uneven distribution of organisms across resources with that of
observational visits across places. Our evidence about observers
consists of recorded checklists, not a census of individual people;
the comparison concerns spatial concentration, not identification
of ecological abundance from volunteer records.

Modern platforms such as eBird can extend ecological monitoring
far beyond a designed field survey. Yet birdwatchers choose
where and when to look, making the spatial distribution of
checklists highly uneven \citep{KellingEtAl2019,JohnstonEtAl2020}.
A complete checklist records detected species during a visit and
provides a geographic sampling event; detection probability can
still differ among species.

The common language is a density ratio. For population ecology
it compares local abundance with the resources available at a site.
For participatory monitoring the formal counterpart compares the
location distribution of records with an ecological reference;
that reference cannot be identified from checklist locations alone.
In the numerical examples, the geographic reference is stipulated
by design; the eBird illustration reports recording locations only.
Hill numbers \citep{hill1973diversity,chao2014unifying}
summarize concentration relative to either reference, and the same reference-measure formulation provides a common algebra.
The corresponding ecological questions concern how many
species are effectively represented and how much space is effectively
surveyed.
The information geometry of maximum Hill diversity distributions
has been investigated by \citet{eguchi2025maxdiv}. 

Our contribution has three parts. First, for generalized logistic
regulation with conservative migration we identify a Hill-number
effective carrying capacity, including a space--time identity for
periodically varying environmental resources. The result quantifies
crowding mismatch, while related upper bounds are already known \citep{guo2020capacity}.
Second, we show how fitness-driven movement provides a concrete
mechanism selecting the same Hill order for demographic regulation
and porous-medium transport. We formulate its free energy in Hill-number
terms, while distinguishing this interpretation from existing
Hellinger--Kantorovich gradient-flow theory
\citep{kondratyev2016fitness,kondratyev2020reaction}.
Third, we distinguish biological distributions from checklist visits,
identify the effect of shared recording effort on regional summaries,
and give the exact geographic transport limitation imposed by unvisited
locations. Synthetic examples and a deliberately descriptive eBird
illustration document what these indices do, and do not, identify.

Our illustration concerns Watarase-yusuichi, a wetland and
flood-control landscape in Japan's Kanto region. Its history
includes the Ashio Copper Mine pollution, resistance by
Shozo Tanaka and residents of Yanaka village, and later
conservation of village sites during the construction of
Lake Yanaka \citep{mlitWataraseHistory,tochigiWataraseHistory}.
This civic history makes the wetland an appropriate setting
for studying patterns of volunteer attention. It does not,
without a spatial overlay and independent evidence, establish
that any particular recording hotspot was caused by that
history.

Section~\ref{sec:hill_pme} introduces Hill numbers relative to reference
measures. Section~\ref{sec:observation} establishes their links to demographic
regulation, spatial spread, and observer effort. Section~\ref{sec:watarase} presents reproducible numerical examples
and a brief checklist-location illustration at Watarase.
Section~\ref{sec:democratic_ecology} discusses ecological interpretation and survey design.
\section{Framework: species, space, and resources}
\label{sec:hill_pme}

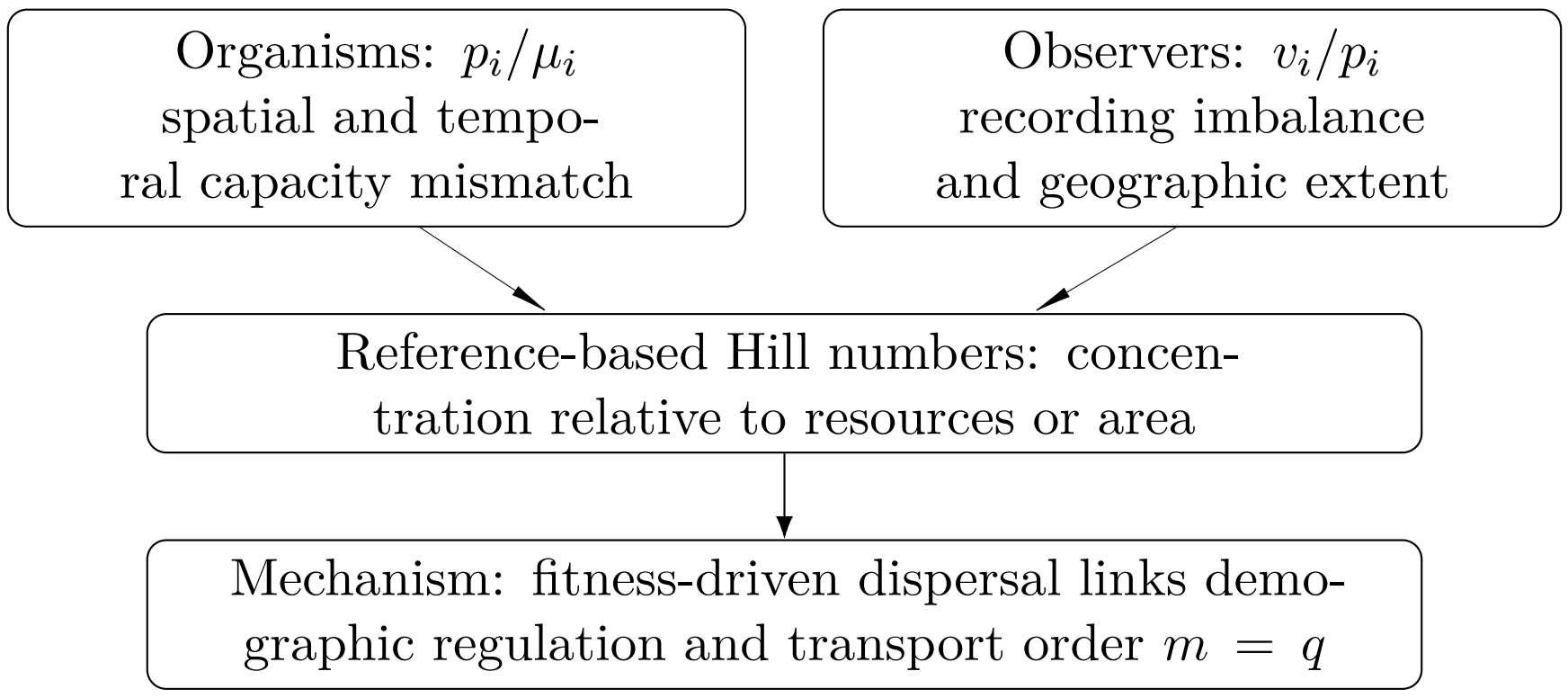
\begin{figure}[t]
\centering
\begin{tikzpicture}[>=Latex,
 layer/.style={draw,rounded corners,align=center,font=\small,
 text width=5.15cm,minimum height=1.03cm},
 bridge/.style={draw,rounded corners,align=center,font=\small,
 text width=9.1cm,minimum height=0.9cm}]
 \node[layer] (eco) at (-3.0,1.5)
 {Organisms: $p_i/\mu_i$\\spatial and temporal capacity mismatch};
 \node[layer] (eff) at (3.0,1.5)
 {Observers: $v_i/p_i$\\recording imbalance and geographic extent};
 \node[bridge] (common) at (0,-0.45)
 {Reference-based Hill numbers: concentration relative to resources or area};
 \draw[-{Latex[length=3mm,width=.8mm]},line width=.2pt]
   ([xshift=3.2mm]eco.south) -- ([xshift=-17.3mm]common.north);
 \draw[-{Latex[length=3mm,width=.8mm]},line width=.2pt]
   ([xshift=-1.2mm]eff.south) -- ([xshift=18.2mm]common.north);
 \node[bridge] (link) at (0,-2.15)
 {Mechanism: fitness-driven dispersal links demographic regulation and transport order $m=q$};
 \draw[->] (common.south) -- (link.north);
\end{tikzpicture}
\caption{Reference-measure Hill numbers summarize two distinct
spatial imbalances. On the organism side, conservative movement and
crowding yield a space--time effective-capacity identity; under
fitness-driven movement the same Hill order enters nonlinear
transport. On the observation side, geographic concentration
is measurable from visits, while the ecological reference is
not identifiable from locations alone.}
\label{fig:two_layer_framework}
\end{figure}

An ecological community is described both by \emph{what} species it contains
and by \emph{where} its members occur. A species inventory alone cannot tell
whether the same population is concentrated in a few refuges or dispersed
across a large landscape. We first recall how Hill numbers express an
effective number of species, then apply the same idea to occupied space.
This leads naturally to a question about dynamics: how does effective
occupancy change when organisms move away from crowded patches?
Figure~\ref{fig:two_layer_framework} summarizes the two density ratios
studied in this paper and the reference-based diversity measure they share.

Hill numbers were introduced as an effective number of species
\citep{hill1973diversity,chao2014unifying}. For relative species
abundances $p(s)$, $s=1,\ldots,S$, with $\sum_s p(s)=1$, the
order-$q$ Hill number is
\begin{equation}
{}^q\!D_{\mathcal S}
=\left(\sum_{s=1}^S p(s)^q\right)^{1/(1-q)},
\quad q>0,\quad q\ne1.
\label{eq:hill_species}
\end{equation}
For $S$ equally abundant species it equals $S$. Its order-zero
extension is species richness, its order-one limit is the exponential
of Shannon entropy, and its order-two value is the inverse Simpson
index \citep{jost2006entropy}. Increasing $q$ emphasizes more abundant
species.

The same principle describes how widely individuals occupy geographic
space. Let $p(x)$ be a probability density on
$\Omega\subset\mathbb R^d$, with $\int_\Omega p(x)\,dx=1$. Its
\emph{spatial Hill number} is
\begin{equation}
{}^q\!D_\Omega(p)
=\left(\int_\Omega p(x)^q\,dx\right)^{1/(1-q)}.
\label{eq:hill_spatial}
\end{equation}
This is the exponential of the continuous R\'enyi entropy and is
related to the measure of distributional extent of
\citet{campbell1966extent}; see also the spatial-information review of
\citet{angulo2021spatial}. Unlike \eqref{eq:hill_species},
\eqref{eq:hill_spatial} has units of area in two dimensions and volume
in three. If $p$ is uniform on a region $A\subseteq\Omega$ of finite
positive volume, then ${}^q\!D_\Omega(p)=|A|$ for every $q$. Thus
``effective number of species'' becomes ``effective occupied area''.
In practice these areas are defined relative to a fixed geographic
reference and spatial resolution.

For heterogeneous resources, species and space are unified by a
reference-measure definition. Let $P$ be a probability measure
absolutely continuous with respect to a positive $\sigma$-finite
measure $\nu$ on $\mathcal X$. Define
\begin{equation}
{}^q\!D_\nu(P)
=\left\{\int_{\mathcal X}\left(\frac{dP}{d\nu}\right)^q
\,d\nu\right\}^{1/(1-q)},\quad q>0,\quad q\ne1.
\label{eq:relative_hill}
\end{equation}
Counting measure on species recovers \eqref{eq:hill_species};
Lebesgue measure recovers \eqref{eq:hill_spatial}. The order-one
limit is ${}^1\!D_\nu(P)=\exp(-\int\log(dP/d\nu)\,dP)$,
whenever finite, and the order-zero extension is the reference
measure of the support, when finite. The homogeneity relation
${}^q\!D_{c\nu}(P)=c\,{}^q\!D_\nu(P)$ holds for $c>0$.
If $0<\nu(\mathcal X)<\infty$, Jensen's inequality yields
${}^q\!D_\nu(P)\leq\nu(\mathcal X)$, with equality exactly when
$P=\nu/\nu(\mathcal X)$. With a probability reference measure,
$-\log {}^q\!D_\nu(P)$ is the R\'enyi divergence of order $q$
\citep{renyi1961entropy}. More generally, when $\nu(\mathcal X)<\infty$,
the Hill number is a monotone transform of Amari's $\alpha$-divergence
between the positive measures $P$ and $\nu$ with $\alpha=1-2q$:
\[
D_\alpha(P\,\|\,\nu)
=\frac{1}{q(1-q)}\int_{\mathcal X}
\left\{q\,\frac{dP}{d\nu}+1-q-\left(\frac{dP}{d\nu}\right)^q\right\}d\nu
\]
Hence, a larger Hill number corresponds
to a smaller $\alpha$-divergence from the reference, and the limit
$q\to1$ corresponds to $\alpha=-1$, the generalized Kullback--Leibler case.
The $\alpha$-divergences induce the dual $\alpha$-connections of
information geometry
\citep{amari2000methods,amari2016information}.
Related diversity partitioning is considered by
\citet{reeve2016partition}; the information geometry of
maximum-diversity distributions by \citet{eguchi2025maxdiv}.

For a spatial grid with $J$ equal-area cells of area
$a=|\Omega|/J$, \eqref{eq:relative_hill} uses the masses $\nu_i=a$.
For $J$ site capacities $k_i>0$, the choice $\nu_i=k_i$ instead
gives ${}^q\!D_k(p)$, whose units are individuals. This last choice,
not a new definition of diversity, is what turns the Hill number into
the effective capacity in Section~\ref{subsec:gadgil}.

For multispecies communities, let $p(s,x)$ denote a joint
species--location probability density, with spatial marginal
$p(x)=\sum_s p(s,x)$. Its local conditional composition is
$p(s\mid x)=p(s,x)/p(x)$ where $p(x)>0$.
The spatial Hill number of $p(x)$ measures the geographic
concentration of the community without introducing an additional
joint diversity index that is not used below.

The spatial index also connects to familiar measures of
aggregation. When $\Omega$ is partitioned into $J$ equal-area cells,
with counts $n_i$ and total $N=\sum_{i=1}^J n_i>1$,
Lloyd's mean crowding is
$m_{\mathrm L}=\sum_i n_i(n_i-1)/N$. The piecewise-constant density is
$p(x)=Jn_i/(|\Omega|N)$ on cell $i$, giving
${}^2\!D_\Omega=(|\Omega|/J)N^2/\sum_i n_i^2$. Since
$\sum_i n_i^2=N(1+m_{\mathrm L})$,
\[
{}^2\!D_\Omega
=\frac{|\Omega|}{J}\frac{N}{1+m_{\mathrm L}}.
\]
Morisita's dispersion index is
$I_\delta=Jm_{\mathrm L}/(N-1)$. Since $\sum_in_i(n_i-1)/(N(N-1))$
is unbiased for the probability that two records fall in the same cell
when $N$ records are allocated to cells independently, $|\Omega|/I_\delta$ is
the order-two effective area computed by inverting this unbiased
concentration estimate, provided it is positive; the resulting area
index is not itself unbiased and is close to ${}^2\!D_\Omega$
when the number of records is large relative to the effective number
of occupied cells.
Thus order-two spatial Hill diversity
is an effective-area counterpart of classical crowding and
aggregation indices \citep{lloyd1967crowding,morisita1959dispersion}.

The same reference-measure definition applies to occupancy relative to
resource availability. For local capacities $k_i>0$ with $K=\sum_i k_i$, let
$\mu_i=k_i/K$ be the normalized resource distribution.
For a probability vector $p$, equation~\eqref{eq:relative_hill}
gives
\[
{}^q\!D_k(p)=K\,{}^q\!D_\mu(p),\quad
{}^q\!D_k(p)\leq K,\quad {}^q\!D_\mu(p)\leq1.
\]
The quantity ${}^q\!D_k(p)$ has units of individuals and will
represent an \emph{effective carrying capacity}; ${}^q\!D_\mu(p)$
is its dimensionless fraction of total capacity. Equality holds
precisely when $p_i=\mu_i$ at every site. For $J$ equal-area cells,
$\mu_i=1/J$ implies ${}^q\!D_\mu(p)={}^q\!D_\Omega(p)/|\Omega|$.
The Hill order describes concentration relative to the chosen
reference; it does not specify a movement mechanism.

\section{Main results: ecological crowding and observation effort}
\label{sec:observation}

\subsection{Gadgil's population bound as a Hill identity}
\label{subsec:gadgil}
The equilibrium population of a fragmented landscape depends on
how its individuals are distributed among local carrying capacities.
Consider a single species with abundance $n_i(t)$ at site $i$,
local capacity $k_i>0$, and movement flux $T_i(n)$ that satisfies $\sum_{i=1}^JT_i(n)=0$ for all $n\in\mathbb R_+^J$. The model is
\begin{equation}
\frac{dn_i(t)}{dt}
=g n_i(t)\left\{1-\left(\frac{n_i(t)}{k_i}\right)^{q-1}\right\}
+T_i(n(t))
\label{eq:general_logistic}
\end{equation}
where $g>0$, $q>1$,  and  $n(t)=(n_i(t))_{i=1}^J$.
The local growth term follows the generalized logistic
model of \citet{gilpin1973global}, while the coupling of
local population regulation and dispersal is standard
in multi-patch population models
\citep{arditi2015dispersal}.
It allows nonlinear, asymmetric, or density-dependent movement:
only the absence of mortality or immigration \emph{during movement}
is imposed. We assume that the migration map $T=(T_i)_i$ is locally Lipschitz and that
$T_i(n)\geq0$ whenever $n_i=0$, so that abundances remain nonnegative.
The intrinsic growth rate $g$ and the crowding
order $q$ are common to all sites; the density dependence has exponent $q-1$. Write
$N=\sum_i n_i>0$, $p_i=n_i/N$, $K=\sum_i k_i$, and $\mu_i=k_i/K$.
Equilibrium claims below concern positive equilibria, when they exist.

For time-dependent capacities, replace $k_i$ by positive, continuous,
$\tau$-periodic functions $k_i(t)$ in \eqref{eq:general_logistic},
and allow a conservative time-dependent flux. Write
${}^q\!D_{k(t)}(p)$ for the Hill number with reference masses
$k_i(t)$. For a positive $\tau$-periodic solution, set
$\bar N=\tau^{-1}\int_0^\tau N(t)\,dt$ and
$\bar K=\tau^{-1}\int_0^\tau\sum_i k_i(t)\,dt$, and define measures
on the site--time space by
\[
P(i,dt)=\frac{n_i(t)}{\tau\bar N}\,dt,
\quad \nu(i,dt)=\frac{k_i(t)}{\tau}\,dt.
\]

\begin{theorem}[Spatiotemporal Crowding--Hill identity]
\label{thm:spatiotemporal_hill}
Under the preceding assumptions, the total abundance satisfies
\begin{equation}
\frac{dN}{dt}=gN\left\{1-
\left(\frac{N}{{}^q\!D_{k(t)}(p)}\right)^{q-1}\right\}.
\label{eq:gadgil_hill_dynamic}
\end{equation}
For a positive $\tau$-periodic solution,
\begin{equation}
\bar N={}^q\!D_\nu(P)\le\bar K.
\label{eq:spatiotemporal_hill}
\end{equation}
\end{theorem}
\begin{proof}
Summing the population equations gives
\[
\frac{dN}{dt}
=
gN-g\sum_i n_i^q k_i(t)^{1-q}.
\]
Since
\[
\sum_i n_i^q k_i(t)^{1-q}
=
N^q\{{}^q\!D_{k(t)}(p)\}^{1-q},
\]
the first identity follows.

For a periodic solution, integration over one period
gives
\[
\frac1\tau\int_0^\tau
\sum_i n_i(t)^q k_i(t)^{1-q}\,dt
=\bar N.
\]
Consequently,
\[
\{{}^q\!D_\nu(P)\}^{1-q}
=
\bar N^{-q}
\frac1\tau\int_0^\tau
\sum_i n_i(t)^q k_i(t)^{1-q}\,dt
=
\bar N^{1-q}.
\]
Finally, Jensen's inequality, applied to the
probability measure $\nu/\bar K$, gives
${}^q\!D_\nu(P)\le\bar K$.
\end{proof}
Equality in the bound requires $P=\nu/\bar K$, hence
$n_i(t)/k_i(t)=\bar N/\bar K$. The integrated balance then gives
$\bar N=\bar K$, so $n_i(t)=k_i(t)$ almost everywhere.
For a nonperiodic interval $[0,T]$, the integrated mass balance
contains the boundary term $(N(T)-N(0))/(gT)$; the displayed
space--time identity should not be applied without controlling
that term. Under bounded, persistent dynamics it gives a
long-time average identity whenever the relevant averages converge.

At a positive equilibrium $n^*=(n_i^*)_{i=1}^J$, write
$N^*=\sum_i n_i^*>0$ and $p_i^*=n_i^*/N^*$.
Then $dN/dt=0$ in \eqref{eq:gadgil_hill_dynamic}, and hence
\begin{equation}
N^*={}^q\!D_k(p^*)\le K.
\label{eq:gadgil_hill_stationary}
\end{equation}
The inequality follows from Jensen's inequality applied
to $\mu_i=k_i/K$, with equality if and only if
$p_i^*=\mu_i$ for all $i$.
This identifies the precise capacity shortfall rather than
reproving an upper bound already known in related logistic
population models \citep{guo2020capacity}. It does not assert
the existence or uniqueness of a positive equilibrium.

The same identity extends to site-specific growth rates.
Replace $g$ in \eqref{eq:general_logistic} by constants
$g_i>0$, retaining a common $q>1$ and conservative
migration. At a positive equilibrium, let
$B^*=\sum_i g_i n_i^*>0$ and
$b_i^*=g_i n_i^*/B^*$.
Summing the equilibrium equations gives
$B^*=\sum_i g_i(n_i^*)^q k_i^{1-q}$,
and hence
$\sum_i(b_i^*)^q(g_i k_i)^{1-q}=(B^*)^{1-q}$.
Consequently, Jensen's inequality yields
\begin{equation}
B^*={}^q\!D_{gk}(b^*)\le\sum_i g_i k_i.
\label{eq:growth_weighted_hill}
\end{equation}
The bound in \eqref{eq:growth_weighted_hill} applies to
growth-weighted abundance, not necessarily to $N^*$ itself;
the latter can exceed $K$ when growth rates differ.

To quantify the spatial mismatch between population abundance
and resource availability, consider the relative density
$p_i/\mu_i$. Its resource-weighted mean is one, since
$\sum_i\mu_i(p_i/\mu_i)=1$. Define its squared coefficient
of variation by
\[
\operatorname{CV}_\mu(p/\mu)^2
=\sum_i\mu_i\left(\frac{p_i}{\mu_i}-1\right)^2
=\sum_i\frac{(p_i-\mu_i)^2}{\mu_i}.
\]
This quantity vanishes precisely when $p=\mu$ and measures
spatial heterogeneity in population abundance relative to
resource availability. For the ordinary logistic law $q=2$,
the stationary formula therefore becomes
\begin{equation}
\frac{N^*}{K}
={}^2\!D_\mu(p^*)
=\frac{1}{1+\operatorname{CV}_\mu(p^*/\mu)^2}.
\label{eq:gadgil_cv}
\end{equation}
Equivalently, $N^*={}^2\!D_k(p^*)$. At equilibrium,
$p_i^*/\mu_i=(K/N^*)(n_i^*/k_i)$, so this coefficient
of variation measures heterogeneity in local crowding
$n_i^*/k_i$ under capacity weights.
This quantifies the spatial crowding imbalance central to
\citet{gadgil1971dispersal}, with its Hill order fixed by
the demographic regulation law. In the logistic case, if
$c_i=n_i^*/k_i$ and $\bar c=\sum_i\mu_i c_i=N^*/K$,
then \eqref{eq:gadgil_cv} is equivalent to
$N^*/K=1-\operatorname{Var}_\mu(c)/\bar c$.
Thus the classical crowding-variance interpretation and
the reference-based Hill representation are identical statements
of demographic balance. The identity is a resource-weighted expression of aggregate
logistic balance. Whether
dispersal among heterogeneous patches raises or lowers total abundance
has a long history. Dispersal can raise total equilibrium abundance
above the sum of local capacities
\citep{freedman1977mathematical,holt1985population}, and the outcome
depends on how intrinsic growth rates covary with capacities
\citep{lou2006migration,arditi2015dispersal,deangelis2016dispersal,guo2020capacity};
this has also been tested experimentally with yeast populations
\citep{zhang2017carrying}. For common growth rates, the upper bound is known in related
logistic dispersal models \citep{guo2020capacity}.
Equation~\eqref{eq:gadgil_hill_stationary} identifies its exact
capacity shortfall as a Hill-number mismatch for every conservative
migration flux. The only lossless configuration
is the ideal free distribution of \citet{fretwell1969territorial}.
Unconditional dispersal is selected against in environments that vary
in space but not in time \citep{hastings1983can}, whereas dispersal
strategies that produce the ideal free distribution are evolutionarily
stable \citep{cantrell2010evolution}. The identity therefore quantifies
the demographic cost of departures from ideal free dispersal, the
question of dispersal evolution that \citet{gadgil1971dispersal}
addressed. The regulation law \eqref{eq:general_logistic} is the
theta-logistic model of \citet{gilpin1973global}, and concave
regulation, with exponent below one, is common in natural populations
\citep{sibly2005regulation}.

The squared coefficient of variation introduced above
is also Pearson's chi-square divergence. More generally, by
\eqref{eq:gadgil_hill_stationary}, the loss relative to
aggregate capacity is exactly $1-{}^q\!D_\mu(p^*)$.
This is a maximum-diversity characterization
\emph{relative to resources}. It relates to the variational
characterization of maximum-diversity distributions
\citep{eguchi2025maxdiv}, although optimization under
species-level constraints and capacity-weighted occupancy
are different problems.

\begin{remark}[Which exponent is selected?]
\label{rem:exponent_selected}
The demographic crowding law in \eqref{eq:general_logistic}
selects Hill order $q$, whereas the migration mechanism
changes the realized proportions $p_i^*$.
With growth scaled by $g/(q-1)$ and $q\downarrow1$,
the per-capita rate approaches $-g\log(n_i/k_i)$
(Gompertz regulation); the stationary identity becomes
$N^*={}^1\!D_k(p^*)$,
the order-one limit of \eqref{eq:relative_hill}.
A particular fitness-driven movement rule can select the same
order $q$ for transport; see Theorem~\ref{thm:fitness_hill}.
\end{remark}

At a positive equilibrium, births occur at site $i$ at rate $gn_i^*$
and crowding deaths at rate $gn_i^*(n_i^*/k_i)^{q-1}$. The normalized spatial distribution of births
is $p^*$, and that of crowding deaths is
\[
d_i^*=\frac{(p_i^*)^qk_i^{1-q}}
{\sum_j(p_j^*)^qk_j^{1-q}}.
\]

\begin{proposition}[Births, crowding deaths and resources]
\label{prop:escort_duality}
At every positive equilibrium of \eqref{eq:general_logistic},
\[
N^*={}^q\!D_k(p^*)={}^{1/q}\!D_k(d^*)\le K.
\]
Moreover,
\[
T_i(n^*)=gN^*(d_i^*-p_i^*).
\]
For $p_i^*>0$, $d_i^*/p_i^*=(n_i^*/k_i)^{q-1}$.
\end{proposition}
\begin{proof}
By the escort definition,
\[
\sum_i(d_i^*)^{1/q}k_i^{1-1/q}
=\left(\sum_j(p_j^*)^qk_j^{1-q}\right)^{-1/q}.
\]
Raising both sides to $q/(q-1)$ gives
${}^{1/q}\!D_k(d^*)={}^q\!D_k(p^*)$.
The equality to $N^*$ and its upper bound follow from
\eqref{eq:gadgil_hill_stationary}. At equilibrium, the total
crowding-death rate equals $gN^*$, so deaths at site $i$ occur at
rate $gN^*d_i^*$; balancing births, deaths and movement proves
the flux identity. Dividing the death rate by $gn_i^*$ gives
the stated ratio for $p_i^*>0$.
\end{proof}

Proposition~\ref{prop:escort_duality} identifies the spatial
crowding-death distribution with the escort of births relative
to carrying capacity. The same effective capacity is obtained
at order $q$ from births and at order $1/q$ from deaths;
all three spatial distributions agree at the ideal free
configuration. The flux identity also shows that, at a site with
$p_i^*>0$, net inward movement ($T_i(n^*)>0$) is equivalent
to $n_i^*>k_i$: inflow balances crowding deaths in excess of
local births. This is a consequence of the common-growth,
conservative-movement assumptions, not a claim about every
migration mechanism. The raw counts and capacities can be used directly
after normalization to $p_i=n_i/N$. In particular, if capacities
are known up to a common factor, an equilibrium census and
\eqref{eq:gadgil_hill_stationary} can determine that factor.

For example, take two sites with $q=2$, $g=1$ and
$k_1=k_2=100$. The distribution $n^*=(120,40)$ has
$N^*=160={}^2\!D_k(0.75,0.25)$. Local growth contributions
are $-24$ and $+24$, respectively, so equilibrium requires
$T(n^*)=(+24,-24)$. The normalized crowding-death distribution is
$d^*=(0.9,0.1)$; the reciprocal-order identity is
${}^{1/2}\!D_k(d^*)=(\sqrt{90}+\sqrt{10})^2=160$.
Thus the over-capacity first site receives net inflow: its local crowding
loss is balanced by immigrants from the less crowded second site.

\subsection{Fitness-driven dispersal and common Hill order}
\label{subsec:fitness}
In Section \ref{subsec:gadgil}, the conservative migration flux $T_i$ in \eqref{eq:general_logistic} is left unspecified, so the demographic Hill order need not coincide with the exponent governing dispersal. 
We now consider a particular mechanism that links these two processes: individuals move up gradients of local per-capita fitness, which is determined by the same density-dependent regulation law. When carrying capacity is spatially constant, this mechanism produces porous-medium diffusion with order $m=q$.
Fitness-dependent dispersal has been studied
in ecological models of the ideal free distribution
\citep{cosner2005,cantrell2008},
following the classical framework of
\citet{fretwell1969territorial}.
Related gradient-flow formulations have been
developed by
\citet{kondratyev2016fitness,kondratyev2020reaction}.

We adopt the population-density mobility $u$,
corresponding to a discrete migration flux
whose mobility is proportional to local abundance.
This choice is essential: with a different mobility,
fitness-driven transport need not select the same
exponent as demographic regulation.
Consider the fitness-driven dispersal equation
\begin{equation}
\partial_t u=-\nabla\cdot(u\nabla f)+uf,
\quad
f=g\left\{1-\left(\frac{u}{k}\right)^{q-1}\right\}.
\label{eq:fitness_pde}
\end{equation}
Here $\Omega\subset\mathbb R^d$ is a smooth bounded domain,
$g>0$, $q>1$, and $k\in C^2(\overline\Omega)$ is
strictly positive. Let $u>0$ be a sufficiently smooth
solution satisfying $u\nabla f\cdot\mathbf n=0$
on $\partial\Omega$. Set $N=\int_\Omega u\,dx$,
$p=u/N$, and
\begin{equation*}
\mathcal E_q(u)
=\frac{g}{q}\int_\Omega u^qk^{1-q}\,dx-gN.
\end{equation*}
By the definition of the Hill number,
\begin{equation*}
\mathcal E_q(u)
=gN\left\{\frac1q
\left(\frac{N}{{}^q\!D_k(p)}\right)^{q-1}-1\right\}.
\end{equation*}

\begin{theorem}[Hill mass balance and dissipation]
\label{thm:fitness_hill}
In the setting above,
\begin{equation}
\frac{dN}{dt}
=gN\left\{1-
\left(\frac{N}{{}^q\!D_k(p)}\right)^{q-1}\right\},
\label{eq:fitness_mass}
\end{equation}
and
\begin{equation}
\frac{d\mathcal E_q}{dt}
=-\int_\Omega u|\nabla f|^2\,dx
-\int_\Omega uf^2\,dx\le0.
\label{eq:fitness_dissipation}
\end{equation}
\end{theorem}

\begin{proof}
Integrating \eqref{eq:fitness_pde} over $\Omega$
and using the no-flux boundary condition gives
\[
\frac{dN}{dt}
=\int_\Omega uf\,dx
=gN-g\int_\Omega u^q k^{1-q}\,dx
=gN-gN^q({}^q\!D_k(p))^{1-q},
\]
which proves \eqref{eq:fitness_mass}.

The first variation of the energy is
\[
\frac{\delta\mathcal E_q}{\delta u}
=g\left(\frac{u}{k}\right)^{q-1}-g=-f.
\]
Consequently, differentiation along
\eqref{eq:fitness_pde} and integration by parts yield
\[
\begin{aligned}
\frac{d\mathcal E_q}{dt}
&=-\int_\Omega f\,\partial_tu\,dx\\
&=\int_\Omega f\nabla\cdot(u\nabla f)\,dx
  -\int_\Omega uf^2\,dx\\
&=-\int_\Omega u|\nabla f|^2\,dx
  -\int_\Omega uf^2\,dx,
\end{aligned}
\]
where the boundary term vanishes by the no-flux
condition. This proves \eqref{eq:fitness_dissipation}.
\end{proof}

For a positive steady state, \eqref{eq:fitness_mass} gives
$N^*={}^q\!D_k(p^*)\le\int_\Omega k(x)\,dx$.
Moreover, \eqref{eq:fitness_dissipation} forces $f=0$, hence
$u^*(x)=k(x)$. This conclusion uses the specified fitness-driven
migration and does not apply to arbitrary conservative movement.
If $k(x)\equiv k$ is constant, direct differentiation yields
\[
\partial_tu=\kappa_q\Delta(u^q)+uf,
\quad\kappa_q=\frac{g(q-1)}{qk^{q-1}}.
\]
Thus this particular mechanism identifies the porous-medium order
$m=q$. For nonconstant $k(x)$, the flux also involves $\nabla k$.
This is the Hill-number representation of a fitness-driven
reaction--transport gradient flow; its general framework is known
\citep{kondratyev2016fitness,kondratyev2020reaction}.

\subsection{Crowding-driven spread in transport geometry}
\label{subsec:pme_ecology}
In an early analysis of dispersal, \citet{gadgil1971dispersal} let the
emigration rate of a site increase as a power of its crowding, with an
exponent that measures how sharply individuals respond to local density,
and argued that spatial differences in carrying capacity favor a sharp
response. The same exponent fixes the continuum limit of the movement
process. Consider a lattice of equal patches in a homogeneous habitat, so
that crowding is proportional to density, and suppose that each individual
leaves its patch at a per-capita rate proportional to the $(m-1)$st power of
the local density and settles in a neighboring patch chosen at random.
The expected flux from a patch to each neighbor is then proportional to
the $m$th power of the density of the source, and the net exchange
between two neighbors is proportional to the difference of these powers.
Formally, under diffusive scaling of space and time, this lattice construction leads to
a porous-medium equation with exponent $m\geq1$: random movement ($m=1$)
gives ordinary diffusion, whereas crowding-sensitive emigration gives
$m>1$. Density-dependent movement of this kind was later used to study the
regulation of inhomogeneous populations and the segregation of competing
species \citep{gurney1975regulation,shigesada1979segregation,shigesada1980dispersing}.

For a single population of density $u(t,x)$ on $\mathbb R^d$, the
porous-medium equation reads
\begin{equation}
\partial_t u=\kappa\Delta(u^m),\quad \kappa>0,\quad m>1,
\label{eq:pme}
\end{equation}
where $\nabla$, $\nabla\cdot$ and $\Delta=\nabla\cdot\nabla$ denote the
gradient, divergence and Laplacian in $x=(x_1,\ldots,x_d)$. The equation
conserves the total mass $M=\int u\,dx$ and, for $m>1$, spreads a
compactly supported population with a finite front speed. It originated in
the physics of porous media \citep{vazquez2007porous}, and related
pressure-driven equations describe cell migration and tissue growth
\citep{perthame2014hele,falco2023tissue}; here it is the continuum
idealization of crowding-sensitive emigration. Its ecological content is
clearest in transport form. With the crowding energy and the velocity of
individuals
\[
\mathcal E_m(u)=\frac{\kappa}{m-1}\int_{\mathbb R^d}u^m\,dx,
\quad
v=-\frac{\kappa m}{m-1}\nabla(u^{m-1}),
\]
equation \eqref{eq:pme} is the continuity equation
$\partial_tu+\nabla\cdot(uv)=0$. The velocity is minus the gradient of the
first variation of $\mathcal E_m$, so transport follows the negative gradient of nonlinear crowding pressure.
Velocity depends on this pressure gradient, not only on local density.

The mass-preserving equation \eqref{eq:pme} is a
$2$-Wasserstein gradient flow \citep{jordan1998variational,otto2001geometry};
its transport kinetic energy is $\int u|v|^2\,dx$
\citep{benamou2000computational}. Here $p_t=u(t,\cdot)/M$ and
${}^q\!D_\Omega(t)$ measures its effective occupied area,
with the Shannon and support-size limits at orders one and zero.

The Zel'dovich--Kompaneets--Barenblatt (ZKB) profile is an explicit
source-type solution of \eqref{eq:pme}:
\begin{equation}
B_M(t,x)=t^{-\alpha}\left(C_M-k_m|x|^2t^{-2\beta}\right)_+^{1/(m-1)}.
\label{eq:zkb}
\end{equation}
Here $\alpha=d/[d(m-1)+2]$, $\beta=1/[d(m-1)+2]$, and
$k_m=\beta(m-1)/(2m\kappa)$.
The constant $C_M>0$ is determined by $\int B_M(t,x)\,dx=M$, and the support
has radius $R(t)=(C_M/k_m)^{1/2}t^\beta$. Every solution of \eqref{eq:pme}
with finite mass approaches the ZKB profile of the same mass as
$t\to\infty$ \citep{vazquez2007porous}, although the profile ceases to be
a solution once growth, environmental drift or interactions between
species are added.

For a probability density $p_t=u(t,\cdot)/M$, the order-$m$
entropy power \citep{savare2014concavity}, expressed through
an effective occupied area, is
\[
\mathcal N_m(t)=({}^m\!D_\Omega(t))^{2/d+m-1}.
\]
Let $B_M$ be the ZKB profile in \eqref{eq:zkb}, normalized to
$p_{B,t}=B_M(t,\cdot)/M$. Denote by $\Lambda_q>0$ its constant
spatial Hill factor and by $c_{\mathrm{ZKB}}>0$ its entropy-power slope.
\begin{proposition}[Self-similar spread and effective area]
\label{prop:zkb_maxent}
For $m>1$ and $t>0$, $p_{B,t}$ uniquely maximizes
${}^m\!D_\Omega$ under its second-moment constraint. Moreover,
\[
{}^q\!D_\Omega(p_{B,t})=\Lambda_q t^{d/[d(m-1)+2]},\quad q\ge0,
\]
and
\[
\mathcal N_m(t)=c_{\mathrm{ZKB}}t.
\]
\end{proposition}
\begin{proof}
Put $\alpha=d/[d(m-1)+2]$ and $\beta=\alpha/d$.
By \eqref{eq:zkb}, $B_M(t,x)=t^{-\alpha}F(xt^{-\beta})$
with $F(y)=(C_M-k_m|y|^2)_+^{1/(m-1)}$.
For the extremal property, let $h=p_{B,t}$ and fix
$\int p\,dx=1$ and $\int |x|^2p\,dx=\int |x|^2h\,dx$.
As $m>1$, maximizing ${}^m\!D_\Omega$ is equivalent to
minimizing $\int p^m\,dx$. On the support of $h$,
$m h^{m-1}=A-B|x|^2$ for suitable $A,B>0$;
outside it, $A-B|x|^2\le0$.
By convexity of $z\mapsto z^m$,
\[
\int(p^m-h^m)\,dx\ge
m\int h^{m-1}(p-h)\,dx\ge0.
\]
Indeed, the two constraints cancel the integral of
$(A-B|x|^2)(p-h)$, while the integrand remaining outside
the support is nonnegative. Strict convexity yields uniqueness.
For any $q>0$, $q\ne1$, changing variables $y=xt^{-\beta}$
gives $\int p_{B,t}^q\,dx\propto t^{-\alpha(q-1)}$.
The same scaling gives the $q=1$ entropy limit and the
$q=0$ support measure. Thus the spatial Hill number scales as
$t^\alpha$. Finally, $\alpha(2/d+m-1)=1$, so
$\mathcal N_m(t)$ is linear in $t$.
\end{proof}
The fixed-moment extremizer and the linear entropy-power scaling
are classical facts. The ecological interpretation
here is an effective occupied area; the diffusion exponent $m$
is independent of a separately specified demographic order $q$,
except in a mechanism such as Theorem~\ref{thm:fitness_hill}.

In Proposition~\ref{prop:spread_hill}, $u$ is a positive classical
solution of \eqref{eq:pme} on $\mathbb R^d$ with mass $M$ and finite
second moment, whose derivatives decay fast enough at infinity to
justify integration by parts and differentiation of the entropy
integrals; the orders $q$ considered are those for which
$Z_q(t)=\int p_t^q\,dx$ and the associated dissipation integrals
are finite. These assumptions concern the pure, potential-free flow. The ZKB profile is
compactly supported and is treated in Proposition~\ref{prop:zkb_maxent}.
We use the classical sharp porous-medium entropy-power inequality
of \citet{savare2014concavity}: for positive, sufficiently smooth
finite-mass solutions with the stated integrability and decay,
$\mathcal N_m(t)$ is concave, and its derivative is at least the
constant slope of the same-mass ZKB solution. The result applies with
diffusivity $\kappa M^{m-1}$ after constant time rescaling.

\begin{proposition}[Effective occupied area under crowding-driven spread]
\label{prop:spread_hill}
For all admissible $q>0$, ${}^q\!D_\Omega(t)$ is nondecreasing.
At order $m$,
\[
{}^m\!D_\Omega(t)=\left(\frac{(m-1)\mathcal E_m(u(t))}{\kappa M^m}\right)^{-1/(m-1)},
\]
and
\begin{equation}
\frac{d}{dt}\log{}^m\!D_\Omega(t)
=\frac{\int u|v|^2\,dx}{(m-1)\mathcal E_m(u(t))}.
\label{eq:area_dissipation}
\end{equation}
Furthermore, $\mathcal N_m$ is concave and
$d\mathcal N_m/dt\ge c_{\mathrm{ZKB}}$.
\end{proposition}
\begin{proof}
Put $Z_q(t)=\int p_t^q\,dx$, so that $\log{}^q\!D_\Omega=\log Z_q/(1-q)$
for $q\neq1$. Since $\partial_tp_t=\kappa M^{m-1}\Delta(p_t^m)$,
integration by parts gives
\[
\frac{dZ_q}{dt}=-q(q-1)m\kappa M^{m-1}\int p_t^{q+m-3}|\nabla p_t|^2\,dx.
\]
The integral is nonnegative, so $Z_q$ is nonincreasing for $q>1$ and
nondecreasing for $0<q<1$, and in both cases $\log Z_q/(1-q)$ is
nondecreasing. For $q=1$ the same computation shows that the Shannon
entropy of $p_t$ has derivative
$m\kappa M^{m-1}\int p_t^{m-2}|\nabla p_t|^2\,dx\geq0$. This proves monotonicity.

For the energy identity, $\mathcal E_m(u)=\kappa M^mZ_m/(m-1)$ gives the first formula.
The velocity is $v=-\kappa mu^{m-2}\nabla u$.
Consequently,
\[
\int u|v|^2\,dx
=\kappa^2m^2M^{2m-1}\int p_t^{2m-3}|\nabla p_t|^2\,dx.
\]
Setting $q=m$ in the derivative of $Z_q$ gives
\[
\frac{dZ_m}{dt}=-\frac{m-1}{\kappa M^m}\int u|v|^2\,dx.
\]
Division by $(1-m)Z_m$ and the identity
$\kappa M^mZ_m=(m-1)\mathcal E_m$ prove
\eqref{eq:area_dissipation}.

Finally, apply the sharp entropy-power inequality stated immediately
before the proposition. The slope of the same-mass ZKB solution is
$c_{\mathrm{ZKB}}$ by Proposition~\ref{prop:zkb_maxent},
which yields both assertions.
\end{proof}

\begin{remark}[Which order does transport select?]
Monotonicity shows that crowding-driven spread enlarges the effective
occupied area at every order, so for the direction of change the order
is only a counting convention. The energy identity singles out one order: the
order-$m$ effective area is a power of the crowding energy that drives
the flow, and its logarithmic growth rate is the kinetic energy of
dispersal divided by $m-1$ times that energy. At any other order the
growth rate is the transport inner product of the gradients of two
different energies and has no such reading. Thus, just as demographic
regulation selects the order $q$ in
\eqref{eq:gadgil_hill_stationary}, the transport geometry of dispersal
selects the order $m$ during spread. By the lattice derivation above,
$m-1$ is the power controlling emigration sensitivity. For the special fitness-driven model of
Theorem~\ref{thm:fitness_hill}, these orders coincide, $m=q$.
\end{remark}

\begin{figure}[t]
\centering
\includegraphics[width=0.98\linewidth]{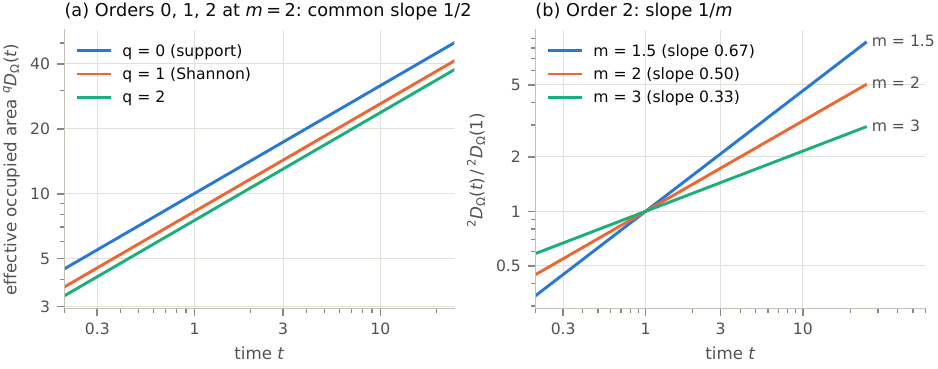}
\caption{The self-similar porous-medium example of
Proposition~\ref{prop:zkb_maxent}, calculated analytically in two
dimensions with unit total mass and diffusivity. (a) Hill orders
$q=0,1,2$ give parallel lines on log--log axes at $m=2$, with exponent
$1/2$. (b) At order $q=2$ the growth exponent is $1/m$; the curves are
scaled to one at $t=1$. In two dimensions $\mathcal N_m$ is the $m$th power of the
order-$m$ effective area, so the order-two curve at $m=2$ corresponds to
linear growth of $\mathcal N_2$. The curves follow directly from the self-similar scaling law.}
\label{fig:zkb_hill_scaling}
\end{figure}

Identity \eqref{eq:area_dissipation} expresses the entropy-production
identity of \citet{savare2014concavity} directly as the growth rate
of order-$m$ effective area. As $m\downarrow1$ it reduces
to de Bruijn's identity for the heat equation, in which the growth rate of
Shannon entropy is the Fisher information \citep{stam1959some}, and the
entropy-power concavity reduces to Costa's concavity of the entropy power
\citep{costa1985new}. The new element here is the reading of these results
as statements about effective occupied area. The concavity result makes that reading
quantitative: at every time, the order-$m$ effective area raised to the
power $2/d+m-1$ grows at least as fast as it does for the self-similar
profile. For admissible solutions, the corresponding entropy power grows at a
rate no smaller than that of the same-mass ZKB solution; its growth
rate is nonincreasing in time. Figure~\ref{fig:zkb_hill_scaling}
illustrates Proposition~\ref{prop:zkb_maxent}; in two dimensions $\mathcal N_m$
is simply the $m$th power of the effective area of order $m$.

Here $t$ denotes physical time. The limit $m\downarrow1$
recovers the heat equation and Shannon entropy, while the formulas
with $(m-1)^{-1}$ apply only to $m>1$. The energy identity in
Proposition~\ref{prop:spread_hill} is for mass-preserving diffusion,
whereas the reaction--transport system of Theorem~\ref{thm:fitness_hill}
uses an unbalanced transport gradient structure.

\subsection{The observational counterpart: concentration of records}
A sighting map is not simply a map of abundance: people visit some
places more often than others. We distinguish (i) biological
individuals or occurrences, (ii) geographic recording or checklist
\emph{visits}, and (iii) species labels or detections reported
at those visits. Let $p(s,x)$ be the ecological joint
species--location distribution of Section~\ref{sec:hill_pme}.
At any location with $p(x)>0$, write $p(s\mid x)=p(s,x)/p(x)$
and define the local Hill number
${}^q\!D_{\mathcal S}(x)=\{\sum_s p(s\mid x)^q\}^{1/(1-q)}$,
with its continuous order-one limit and support-size order-zero extension.
Let $e(x)\geq0$ describe recording effort shared across species.
Under species-independent recording the distribution of records is
\begin{equation}
r(s,x)=\frac{e(x)p(s,x)}{Z_e},\quad
Z_e=\int_\Omega e(x')p(x')\,dx'>0.
\label{eq:observed_thinning}
\end{equation}
An effort field may be known only up to a multiplicative constant;
it need not be a probability. We reserve $\pi(x)$ for an event-level
recording probability. When an ecological occurrence process has
intensity $\lambda_s^{\mathrm{eco}}(x)$, independent thinning
gives $\lambda_s^{\mathrm{obs}}(x)=
\pi(x)\lambda_s^{\mathrm{eco}}(x)$
\citep{WartonShepherd2010,FithianHastie2013}.
When $\pi(x)$ and $\lambda_s^{\mathrm{eco}}(x)$ are positive, this is the
additive decomposition
\[
\log\lambda_s^{\mathrm{obs}}(x)
=\log\lambda_s^{\mathrm{eco}}(x)+\log\pi(x).
\]
It connects to the species-distribution literature
on presence-only Poisson point processes, sampling-bias
correction, and robust divergence estimation
\citep{komori2020quasiPPP,komori2023sdm,saigusa2024robustPPP}.
Environmental covariates may model the first term and
accessibility or observer covariates the second, but
their distinguishability requires assumptions or
independent information; a single presence-only map
cannot identify both unrestricted functions.

Under the common-effort model \eqref{eq:observed_thinning},
the marginal recording density is
$r(x)=e(x)p(x)/Z_e$. Hence, whenever
$e(x)p(x)>0$,
\begin{equation}
r(s\mid x)=p(s\mid x),\quad s=1,\ldots,S.
\label{eq:local_composition}
\end{equation}
Thus, common recording effort leaves local species
composition unchanged, although it alters the spatial
weights used in regional aggregation.

The local Hill numbers of $r(\cdot\mid x)$ and $p(\cdot\mid x)$
therefore coincide at every admissible order. Although common
recording effort preserves species composition
conditional on location, it generally changes spatial weights and
therefore need not preserve regionally aggregated species diversity:
$r(s)=\int r(s,x)\,dx$ need not equal $p(s)$.
Complete checklists report the species that were detected, but do
not guarantee equal detectability. Neither effort nor an ecological
distribution is individually identifiable from their product
without additional information.

\begin{remark}[Invariant composition, effort-dependent estimates]
\label{rem:estimator_invariance}

The identity \eqref{eq:local_composition} holds
at the population level and does not imply exact
equality of estimates based on finitely many visits.
 Suppose that cell $i$
receives $B_i$ independent complete checklists and that species $s$ is
reported on each with probability $\phi_{si}$. The expected number of
species recorded in the cell is $\sum_s(1-(1-\phi_{si})^{B_i})$, which
is nondecreasing in $B_i$ although the composition is unchanged. Plug-in Hill
numbers of incidence shares can also depend on the number of checklists
at every order, although monotonicity is not guaranteed. With a single
nonempty checklist, all orders equal the number of species recorded on it.
Observer effort that leaves local composition invariant therefore still
changes estimated local diversity, through sample completeness. 
\end{remark}

There is also a quantitative analogue of
\eqref{eq:gadgil_cv}. On a discrete grid, let
$p_i$ be the ecological location distribution and
$r_i=e_i p_i/Z_e$ the distribution of recorded locations,
where $Z_e=\sum_i p_i e_i>0$. Define the $p$-weighted coefficient of variation of effort by
\[
\operatorname{CV}_p(e)^2
=\frac{\sum_i p_i(e_i-Z_e)^2}{Z_e^2}.
\]
The concentration of species-level recorded locations relative to the ecological
reference is then determined by effort variation. If the same relation holds
for checklist visits, one may replace $r$ by $v$ throughout the identity.

\begin{proposition}[Observer-effort Hill identity]
\label{prop:effort_hill}
Suppose $p_i>0$ and $e_i>0$ on the relevant support.
For the final assertion, let $n$ locations $X_1,\ldots,X_n$ be sampled
independently from $r$, with inverse-effort weights $1/e_{X_\ell}$ and Kish
effective sample size
$n_{\mathrm{Kish}}=(\sum_\ell e_{X_\ell}^{-1})^2/\sum_\ell e_{X_\ell}^{-2}$. Then
\begin{equation}
{}^2\!D_p(r)=\frac{1}{1+\operatorname{CV}_p(e)^2}
\label{eq:effort_hill}
\end{equation}
and
\begin{equation}
{}^2\!D_r(p)=\frac{\big(\sum_i r_i/e_i\big)^2}{\sum_i r_i/e_i^2}.
\label{eq:inverse_effort_hill}
\end{equation}
Moreover, $n_{\mathrm{Kish}}/n\to{}^2\!D_r(p)$ almost surely.
\end{proposition}
\begin{proof}
By definition
$\sum_i r_i^2/p_i=\sum_i p_i e_i^2/Z_e^2
=1+\operatorname{CV}_p(e)^2$,
which proves \eqref{eq:effort_hill}.
Likewise $p_i=(r_i/e_i)/(\sum_j r_j/e_j)$, giving
\eqref{eq:inverse_effort_hill}. The last assertion
follows by the law of large numbers applied to
$\big(\sum_{\ell=1}^n e_{X_\ell}^{-1}\big)^2/
\sum_{\ell=1}^n e_{X_\ell}^{-2}$
for $X_\ell$ independently distributed according to $r$.
\end{proof}

More generally, for every $q>0$, $q\ne1$, the same Hill definition gives
an exact observer-effort identity:
\[
{}^q\!D_p(r)
=\left\{\frac{\sum_i p_i e_i^q}{Z_e^q}\right\}^{1/(1-q)},
\quad \frac{r_i}{p_i}=\frac{e_i}{Z_e}.
\]
It is the same reference-based Hill construction
as in the demographic equilibrium, but with recording
intensity measured relative to ecological site probability
rather than crowding relative to carrying capacity.
Rescaling the effort field $e$ by any positive constant
leaves $r$ and the identity unchanged. This is an algebraic
statement; observing $r$ does not identify $p$ or $e$
separately.

For checklist sampling, let $v_i$ instead denote the observed
probability of a \emph{visit} to cell $i$, and let $c_{s\mid i}$
denote a stipulated biological conditional species distribution.
An idealized one-label-per-visit model has
$r(s,i)=v_i c_{s\mid i}$, hence $r(s\mid i)=c_{s\mid i}$.
The mechanistic thinning representation $v_i\propto p_i e_i$
requires an ecological reference $p_i$, which is generally
unknown from recorded visits alone. By contrast, the
geographic Hill number of checklist locations is directly
defined on cells: with equal area $a$,
\[
{}^q\!D_\Omega(v)
=a\left(\sum_i v_i^q\right)^{1/(1-q)}.
\]
It must not be confused with ${}^q\!D_p(v)$, which measures
imbalance relative to the \emph{ecological} site distribution.
For example, two unit-area cells with $p=(3/4,1/4)$ and
$v=(1/2,1/2)$ yield ${}^2\!D_p(v)=3/4$, whereas
${}^2\!D_\Omega(v)=2$ unit areas. A complete eBird checklist
can record multiple detected species; a one-label categorical
simulation is an illustration, not a full checklist observation
model.

The forward index in \eqref{eq:effort_hill} is \emph{not}
directly estimable from the sightings alone, since
the ecological reference $p$ is unknown.
The inverse weighting diagnostic, by contrast, can
be evaluated using recorded visits and a \emph{proposed}
positive effort field. It describes the instability
of that reweighting; it does not identify ecological
abundance or establish independent biological coverage.
Thus the ecological and observation layers share
the same algebra, but they have different data and
identification requirements.

\subsection{Geographic support and weighted sample concentration}
\label{subsec:geographic_theory}

Two complementary results distinguish geometric limits of
existing visits from the concentration of their weights.

\begin{proposition}[Sharp geographic support limit]
\label{prop:sharp_support}
Let $\rho=(\rho_1,\ldots,\rho_J)$ be a prespecified
target geographic probability distribution over $J$
cells, and let $V$ be the nonempty set of visited cells.
The 1-Wasserstein distance $W_1(r,\rho)$ with ground transport cost $d(x_i,x_j)$ between cells satisfies
\begin{equation}
\inf_{\operatorname{supp}(r)\subseteq V}
W_1(r,\rho)
=
\sum_{j=1}^J
\rho_j\min_{i\in V}d(x_j,x_i)
=:L_V.
\label{eq:sim_support_bound}
\end{equation}
\end{proposition}

\begin{proof}
Any transport from $\rho$ to a distribution supported
on $V$ incurs a cost of at least the distance from
each target cell $j$ to its nearest cell in $V$.
To attain this bound, assign each mass $\rho_j$ to
a nearest cell in $V$ and let $r$ be the resulting
distribution on $V$.
The transport cost is exactly $L_V$.
\end{proof}

\begin{lemma}[Coarse-graining and Kish effective sample size]
\label{lem:kish_hill}
For nonnegative visit weights $w_\ell$ with $\sum_\ell w_\ell>0$,
set $\omega_\ell=w_\ell/\sum_j w_j$ and
$\widetilde r_i=\sum_{\ell:X_\ell=i}\omega_\ell$.
For equal-area cells of area $a>0$ and every $q>0$,
\[
{}^q\!D(\widetilde r)\le{}^q\!D(\omega).
\]
In particular,
\[
{}^2\!D_\Omega(\widetilde r)
\le a\left(\sum_\ell\omega_\ell^2\right)^{-1}=a\,n_{\rm Kish}.
\]
\end{lemma}
\begin{proof}
For $q>1$, grouping nonnegative weights increases the sum
of $q$th powers; for $0<q<1$, grouping decreases that sum.
Applying the exponent $1/(1-q)$ in either case yields the
inequality. The order-one case follows by continuity.
At $q=2$, $n_{\rm Kish}=1/\sum_\ell\omega_\ell^2$.
\end{proof}
The bound is a finite-sample ceiling and does not, by itself,
explain every observed change in geographic concentration. The
support-limit infimum permits any distribution on $V$; a restricted
family of reweightings need not attain it.

\section{Numerical illustrations: organisms and observers}
\label{sec:watarase}
Three self-contained numerical experiments illustrate the spatial
consequences of Section~\ref{sec:observation} on specified landscapes.
The random seeds, formulas, and numerical procedures are provided
in Appendix~\ref{app:sim_reproduce}. The archive supplies the
compiled figure files and independent code for periodic population
balance, reweighting, optimal transport, and aggregate sensitivity
checks.
The final subsection reports an independent checklist-location summary
and sensitivity calculations on a separately archived sampling subset.

\subsection{Population redistribution and effective carrying capacity}
\label{subsec:sim_demography}
\begin{figure}[htbp]
\centering
\includegraphics[width=0.98\linewidth]{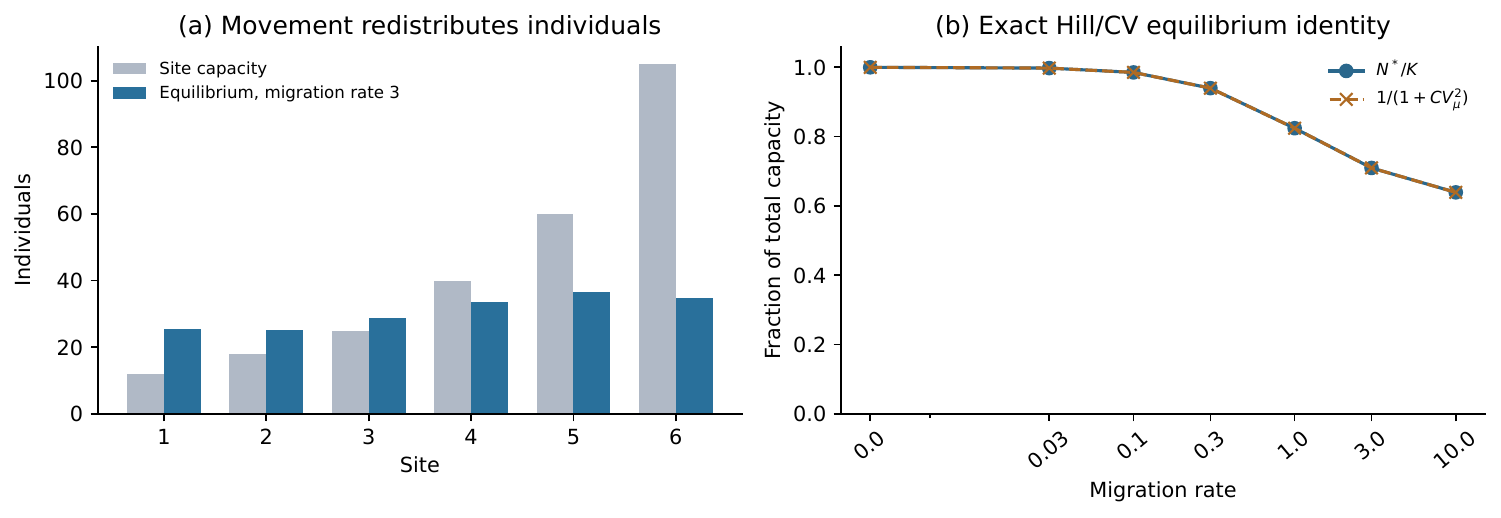}
\caption{Synthetic population redistribution. (a) Local equilibrium
abundances at migration rate $\lambda=3$, relative to local capacities.
(b) Regional capacity utilization agrees with the reciprocal
resource-weighted coefficient-of-variation identity at all migration rates.}
\label{fig:sim_demography}
\end{figure}

Consider six sites connected in a ring with heterogeneous capacities
\[
(k_1,\ldots,k_6)=(12,18,25,40,60,105),\quad K=260.
\]
For ordinary logistic regulation ($g=1$, $q=2$), let symmetric nearest-neighbor
movement have rate $\lambda\geq0$:
\begin{equation}
\dot n_i=n_i(1-n_i/k_i)
 +\lambda(n_{i-1}+n_{i+1}-2n_i),\quad i\ \mathrm{mod}\ 6.
\label{eq:sim_migration}
\end{equation}
Every flux in \eqref{eq:sim_migration} is conservative. Starting with
$n_i(0)=k_i$, we integrate until a numerical equilibrium is reached.
At $\lambda=0$ the ideal resource-matching state $n^*=k$ has
$N^*=K$; migration progressively homogenizes local abundances despite
uneven local resources. Table~\ref{tab:sim_demography} and
Figure~\ref{fig:sim_demography} show the resulting loss of
regional carrying-capacity utilization. In each equilibrium, the
independently computed quantities $N^*/K$ and
$\{1+\operatorname{CV}_\mu(p^*/\mu)^2\}^{-1}$ agree to the numerical
solver's tolerance, as required by \eqref{eq:gadgil_cv}.
This decrease with $\lambda$ is a property of this symmetric ring example,
not a monotonicity theorem for arbitrary dispersal mechanisms.
Theorem~\ref{thm:spatiotemporal_hill} also applies when site
capacities vary periodically; a reproducible illustration follows
Table~\ref{tab:sim_demography}.

\begin{table}[htbp]
\centering\small
\caption{Synthetic logistic population on a six-site ring: equilibrium
capacity utilization and its resource-mismatch coefficient. The two
expressions for utilization agree to numerical precision.}
\label{tab:sim_demography}
\begin{tabular}{@{}rrr@{}}
\toprule
Migration rate $\lambda$ & $N^*/K$ & $\operatorname{CV}_\mu(p^*/\mu)^2$ \\
\midrule
0 & 1.0000 & 0.0000 \\
0.3 & 0.9399 & 0.0640 \\
1 & 0.8243 & 0.2132 \\
3 & 0.7097 & 0.4090 \\
10 & 0.6392 & 0.5644 \\
\bottomrule
\end{tabular}
\end{table}

The space--time identity is illustrated using the same six sites
with $g=1$, $\lambda=0.3$ and seasonal carrying capacities
\[
k_i(t)=k_i\left\{1+0.2\sin\left(t+\frac{2\pi(i-1)}6\right)\right\},
\quad \tau=2\pi.
\]
We numerically integrate for 80 periods and evaluate the last period
as an approximation to the attracting periodic trajectory.
Table~\ref{tab:sim_periodic} checks
\eqref{eq:spatiotemporal_hill} at three demographic orders.
Both averages are independently integrated; the equality is not
imposed in the numerical solver. The values vary with $q$ because
the demographic dynamics, and not merely the diversity convention,
change with the order.
\begin{table}[htbp]
\centering\small
\caption{Synthetic periodic six-site population: time-averaged abundance
and the space--time Hill effective carrying capacity, computed over the
last of 80 periods. All sites have phase-shifted seasonal capacities;
$\bar K=260$ in each case. Values are rounded to six decimals.}
\label{tab:sim_periodic}
\begin{tabular}{@{}rrrr@{}}
\toprule
Hill order $q$ & $\bar N$ & ${}^q\!D_\nu(P)$ & $\bar K$\\
\midrule
1.5 & 234.088272 & 234.088272 & 260.000000\\
2.0 & 242.160783 & 242.160783 & 260.000000\\
3.0 & 248.695163 & 248.695163 & 260.000000\\
\bottomrule
\end{tabular}
\end{table}

\subsection{Shared recording effort and regional species diversity}
\label{subsec:sim_observers}
We next construct a square landscape with $12\times12$ one-square-kilometer
cells. The ecological location reference is uniform, $p_i=1/144$,
as specified by the simulation design. Six artificial species have
location-dependent conditional probabilities $p(s\mid i)$, generated
from Gaussian preferences around six fixed centers. The common
recording-effort field is strictly positive and concentrates visits
around two hotspots (full parameters in Appendix~\ref{app:sim_reproduce}).
Thus
\[
r(s,i)=\frac{p_i e_i p(s\mid i)}{\sum_jp_j e_j},\quad
r(s\mid i)=p(s\mid i),
\]
while the regional species shares generally differ from those under $p$.
We draw 600 independent visit locations from $r_i$ and one artificial
species label per visit from $p(s\mid i)$, with seed 20261114.

\begin{figure}[htbp]
\centering
\includegraphics[width=\linewidth]{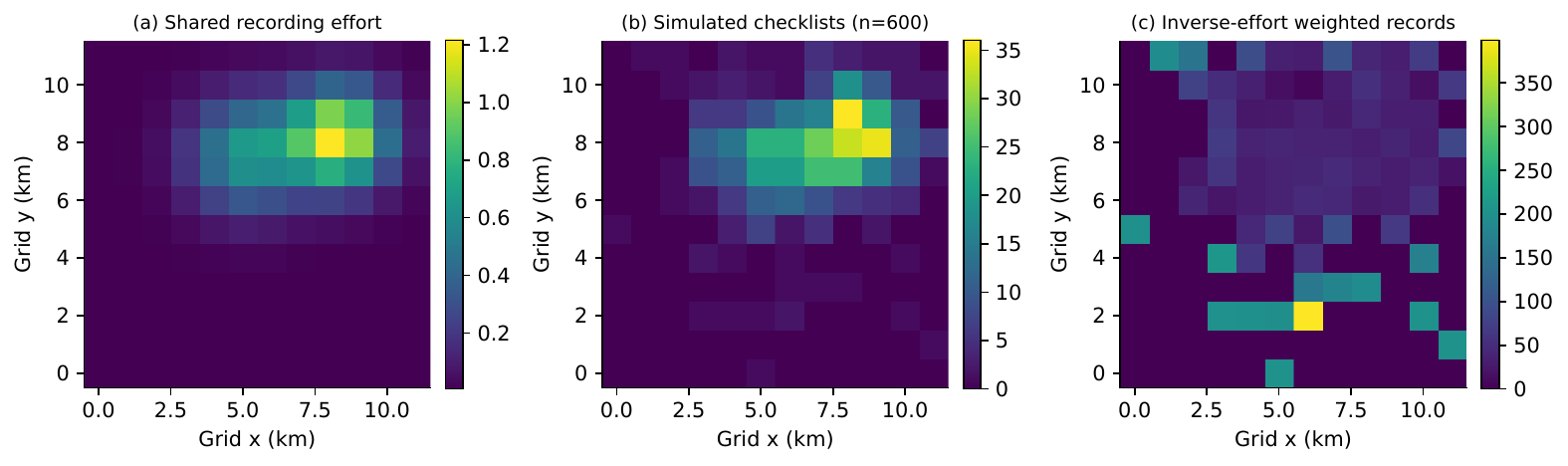}
\caption{Synthetic spatial recording process on 144 one-square-kilometer
cells: (a) the stipulated shared-effort field,
(b) 600 recorded visits, and (c) inverse-effort weighted records.
The recording field is specified by the simulation design. The third panel shows
how rare visits receive large weights; it does not fill unvisited cells.}
\label{fig:sim_observers}
\end{figure}

The stipulated ecological order-two species Hill number is 6.00,
whereas the record-location model has order-two regional species
Hill number 3.22, and the finite sample gives 3.25.
Spatially, the ecological reference occupies all 144 cells uniformly,
but the record-location model has order-two effective area 34.04\,km$^2$
(33.06\,km$^2$ in the sample). Only 75 cells receive a visit,
leaving 69 cells unobserved (Figure~\ref{fig:sim_observers}).
The decrease in regional species diversity does not contradict
\eqref{eq:local_composition}: local species
composition is unchanged, whereas aggregation
over locations uses the biased spatial weights $r_i$.
Both the ecological distribution and recording field are known
parameters of this simulation.

\subsection{Reweighting and the support-limited geographic gap}
\label{subsec:sim_reweight}
Because effort is known in the simulation, we can compute bounded
inverse-effort weights for the recorded site $X_\ell$:
\begin{equation}
w_\ell(c)=\min\{e_{\max}/e_{X_\ell},c\},\quad
\widetilde r_i(c)=
\frac{\sum_{\ell:X_\ell=i}w_\ell(c)}{\sum_{\ell}w_\ell(c)},
\quad c\geq1.
\label{eq:sim_weights}
\end{equation}
For $c=1$ the records are unweighted. We compute Kish's effective 
sample size \citep{kish1965survey} $n_{\rm Kish}(c)=(\sum_\ell w_\ell)^2/\sum_\ell w_\ell^2$
and the exact discrete 1-Wasserstein distance from $\widetilde r(c)$
to the uniform target $u_i=1/144$, using the Euclidean
cell-center distance in kilometers. For the visited set $V=\{i:B_i>0\}$,
Proposition~\ref{prop:sharp_support} gives the exact
support-limited geographic discrepancy.
\begin{table}[htbp]
\centering\small
\caption{Reweighting 600 simulated visits under increasing caps on
inverse-effort weights. Effective surveyed area and regional species Hill
number use order two. $W_1$ refers to the 144-cell uniform geographic
target, not ecological abundance.}
\label{tab:sim_tradeoff}
\begin{tabular}{@{}rrrrr@{}}
\toprule
Cap $c$ & $n_{\rm Kish}/600$ & Area (km$^2$) & Species Hill & $W_1$ (km) \\
\midrule
1 & 1.000 & 33.06 & 3.25 & 3.366 \\
8 & 0.656 & 54.37 & 3.93 & 2.627 \\
32 & 0.320 & 63.53 & 4.53 & 2.067 \\
128 & 0.122 & 49.38 & 5.10 & 0.985 \\
Uncapped & 0.089 & 40.33 & 4.98 & 0.993 \\ 
Nearest-cell optimum$^\dagger$ & 0.068 & 35.63 & 5.35 & 0.676 \\
\bottomrule
\end{tabular}
\par\smallskip
\begin{minipage}{0.92\linewidth}\footnotesize
$^\dagger$Geometric optimum, not an
inverse-effort candidate: each target cell sends its equal mass to
the nearest visited cell, with ties resolved by ascending cell index;
within a visited cell, its assigned mass is divided equally across
checklists. Other tie rules can yield different Kish fractions and
species Hill numbers while leaving the optimal $W_1$ unchanged.
\end{minipage}
\end{table}

\begin{figure}[htbp]
\centering
\includegraphics[width=0.76\linewidth]{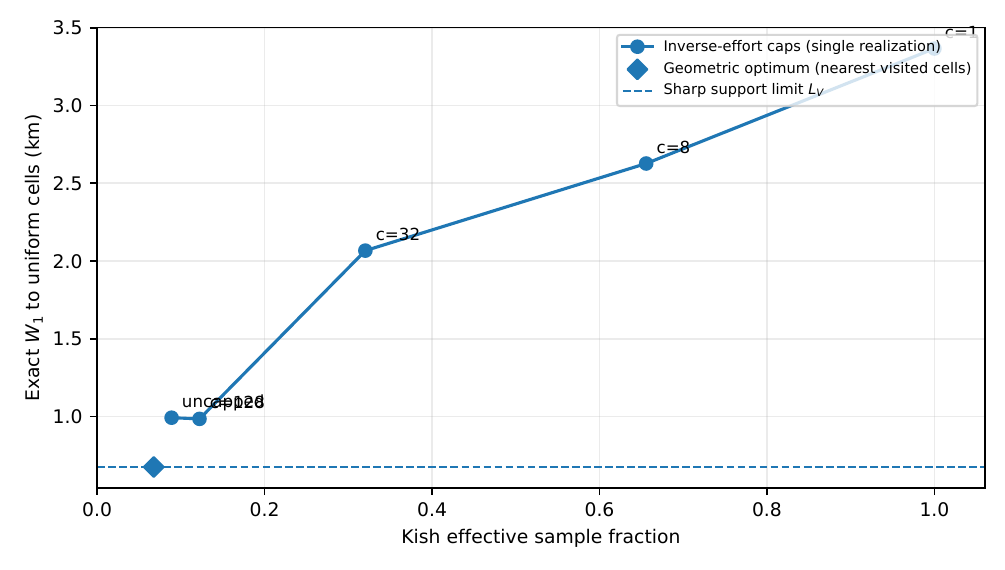}
\caption{Exact 1-Wasserstein distance against Kish effective-sample
fraction for bounded inverse-effort reweighting of simulated visits.
The dashed line is the exact support-limited bound in
\eqref{eq:sim_support_bound}. The diamond shows a nearest-visited-cell allocation attaining
the sharp lower bound; the other points are inverse-effort candidates.}
\label{fig:sim_tradeoff}
\end{figure}

Table~\ref{tab:sim_tradeoff} shows that inverse-effort reweighting
can improve spatial alignment but sharply reduce effective sample size.
The nearest-cell row attains the geometric limit $L_V=0.676371$\,km,
whereas no displayed inverse-effort cap does. The ceiling of
Lemma~\ref{lem:kish_hill} constrains spatial Hill area but does not
fully explain its nonmonotonic finite-sample behavior. In 200 independent
realizations (seed 20261115), the $W_1$ mean (SD) decreases from
3.399 (0.075)\,km without reweighting to 1.148 (0.265)\,km
at cap 128; the corresponding Kish fraction is 0.124 (0.007).
Complete replication summaries are in Table~\ref{tab:repeat_tradeoff}.
No retrospective weighting creates a visit to any of the unvisited cells;
new geographic visits are required to alter the support $V$.

\subsection{A brief Watarase checklist-location illustration}
\label{subsec:sim_watarase}
We use geographic coordinates of eBird sampling events in the
Watarase region, Japan \citep{ebird_ebd,sullivan2009ebird}.
The primary descriptive sample uses the November 2025 sampling-event
release, selecting complete Traveling or Stationary checklists with
positive duration in May--July 2006--2025, within 20\,km of
$(36.2140416^\circ\mathrm{N},139.6731143^\circ\mathrm{E})$.
This taxon-independent extraction yields 617 event identifiers.
On the fixed 1-km grid referenced to that location, 80 cells are visited;
the order-one and order-two effective areas are 25.75 and
13.59\,km$^2$, respectively. The pair-collision-corrected order-two
index is 13.87\,km$^2$. These describe recorded locations, not
independent biological occupancy or the physical area of the wetland.

A separately retained, detailed Watarase extract allows us to
quantify sensitivity to shared checklists and grid resolution.
That archived subset contains 587 complete submissions, of which
177 carry shared-group identifiers representing 55 groups;
collapsing each group to one geographic event reduces its count to
465. Importantly, the archived subset was selected using a fixed
76-label analytical taxon inventory and includes six Historical
checklists, unlike the primary taxon-independent 617-event extraction.
The samples therefore differ in selection and should not be treated
as two versions of the same event population. Table~\ref{tab:watarase_brief}
clearly separates the primary summary from the archived-subset
sensitivity analysis.

\begin{table}[htbp]
\centering\footnotesize
\caption{Watarase geographic concentration and archived-sample sensitivity.
The primary row uses the taxon-independent 617-event extraction. All other rows
are reanalyses of a distinct archived 587-submission subset and must not be
interpreted as sensitivity estimates for the 617-event population.
All areas are equal-cell equivalents in km$^2$.}
\label{tab:watarase_brief}
\begin{tabular}{@{}lrrrrr@{}}
\toprule
Sample and treatment & Cell km & Events & Cells & ${}^1\!D_\Omega$ & ${}^2\!D_\Omega$ \\
\midrule
Primary; not collapsed & 1.0 & 617 & 80 & 25.75 & 13.59 \\
\midrule
Archive; not collapsed & 0.5 & 587 & 94 & 7.38 & 3.54 \\
Archive; group collapsed & 0.5 & 465 & 94 & 7.86 & 3.56 \\
Archive; not collapsed & 1.0 & 587 & 77 & 25.57 & 13.22 \\
Archive; group collapsed & 1.0 & 465 & 77 & 26.59 & 13.06 \\
Archive; not collapsed & 2.0 & 587 & 56 & 79.82 & 48.15 \\
Archive; group collapsed & 2.0 & 465 & 56 & 82.64 & 47.40 \\
\midrule
Archive; through 2024$^\dagger$ & 1.0 & 402 & 66 & 21.99 & 10.58 \\
\bottomrule
\end{tabular}
\par\smallskip
\begin{minipage}{0.92\linewidth}\footnotesize
$^\dagger$The through-2024 row retains
unmerged archived submissions. The primary 617-event sample has not
been regrouped or regridded because its event-level input was not retained in the reproducibility materials; the archived subset provides a separate
sensitivity illustration, not a replacement calculation.
\end{minipage}
\end{table}

Within the archived subset, grouping shared records changes the
order-two area relatively little at each grid width, whereas the
numerical effective area is highly resolution-dependent. A visit
is assigned to the reported checklist coordinate, rather than to
its complete traveled route. Sampling-event identifiers and observer
records are not published with this article; users may obtain the
source eBird Basic Dataset independently under Cornell's access conditions.

\section{Discussion}
\label{sec:democratic_ecology}
The common thread in these models is not a causal equivalence
between organisms and observers, but the use of a reference measure
to quantify spatial concentration. In Gadgil-type population regulation,
local capacities define the resource reference: the logistic equilibrium
loss is exactly the penalty for an abundance distribution that does not
match available resources. Population-conserving movement changes
this distribution, and its demographic effects depend on how migration
redistributes individuals. The symmetric six-site ring illustrates
how greater mixing can lower regional capacity utilization, but
neither the Hill identity nor the simulation claims that every form of
movement has this effect.

The spatiotemporal extension applies the same crowding mismatch
to periodic environmental variation. Demographic order $q$ and
porous-medium exponent $m$ arise independently under generic
movement rules, but fitness-driven dispersal selects $m=q$ in
a homogeneous habitat. Its energy and mass-balance identities
remain valid for heterogeneous resources, even though transport
then includes gradients of $k(x)$. Classical diffusion entropy-power
results describe the transient effective occupied area; the
reaction--transport gradient-flow model provides a concrete
mechanistic bridge between the two descriptions.

For observers, a shared recording field cancels when species composition
is conditioned on location, but it changes the geographic weights
used in regional summaries. In the synthetic example, order-two
species diversity falls from 6.00 under the ecological reference to
3.22 under the record distribution, even though local composition is
unchanged. This is possible because geographic bias and ecological
heterogeneity interact during aggregation. In real citizen-science
data, recording effort and ecological distribution generally cannot
be separated using checklist locations alone. The synthetic field is
known by construction and must not be interpreted as an empirically
identified eBird effort surface.

\emph{Reweighting and additional visits.}
Inverse-effort reweighting improves some representativeness diagnostics
but reduces Kish effective sample size. The synthetic experiment illustrates the sharp minimum possible
distance to the uniform geographic target, determined by the set of
unvisited cells. This separates two
operational questions: how existing records should be weighted and
where future recording should occur. The former cannot solve the
latter. The Watarase checklist summary likewise demonstrates that the number
of visited locations and their Hill-effective area are distinct
descriptors of geographic coverage. Archived-subset sensitivity
results show that effective area depends on grid scale even when
group-shared records are collapsed.

\emph{Democratic ecology and limitations.}
Gadgil's emphasis on local participation encourages monitoring both
who contributes knowledge and which places are represented.
The present calculations address only the latter through recorded
locations. Broader participation does not ensure geographic coverage,
and geographic coverage alone does not ensure equal detection,
ecological accuracy, or social representation. The demographic result
requires common logistic regulation and conservative movement; the
porous-medium example describes a homogeneous idealization, and the
observer-effort model assumes species-independent recording.
Independent ecological surveys and deliberate visits to previously
unrecorded cells are needed to test whether improved geographic
representation also improves biodiversity inference. 

\FloatBarrier
\appendix
\section{Reproducing the simulations and checklist-location summary}
\label{app:sim_reproduce}
The numerical reproduction package includes the following scripts:
\texttt{run\_\allowbreak periodic\_\allowbreak hill.py},
\texttt{run\_\allowbreak reweighting\_\allowbreak checks.py}, and
\texttt{run\_\allowbreak watarase\_\allowbreak sensitivity.py}, with the synthetic
inputs described in Section~\ref{sec:watarase}. The synthetic experiments
use only NumPy, SciPy, pandas, and Matplotlib and require no external datasets.
The random generator is NumPy's default generator with seed 20261114.
The script \texttt{run\_\allowbreak periodic\_\allowbreak hill.py} reproduces
Table~\ref{tab:sim_periodic} with SciPy's BDF solver, integrating
80 periods and applying 4001-point Simpson quadrature on the last
period. The maximum numerical periodic endpoint mismatch in these
runs is less than $2\times10^{-8}$. The script evaluates the two
sides of \eqref{eq:spatiotemporal_hill} independently.
The six-site ring uses capacities $(12,18,25,40,60,105)$ and
$\lambda\in\{0,0.03,0.1,0.3,1,3,10\}$; equilibrium is obtained by
integrating \eqref{eq:sim_migration} from $n(0)=k$ using SciPy's BDF
integrator to time 1500, with relative tolerance $2\times10^{-10}$
and absolute tolerance $2\times10^{-11}$. We check the stationary
residual and both exact Hill identities.

For the observer simulation, integer site coordinates are
$x_i=(a_i,b_i)\in\{0,\ldots,11\}^2$, with $p_i=1/144$.
The cells are indexed with the first coordinate varying fastest, namely
$i=1+a_i+12b_i$.
The common recording field is
\[
\begin{split}
e_i={}&0.006+1.2\exp\!\left(-\frac{\|x_i-(8.3,8.2)\|^2}{2(1.20)^2}\right)\\
&+0.65\exp\!\left(-\frac{\|x_i-(5.1,7.6)\|^2}{2(1.35)^2}\right).
\end{split}
\]
Six artificial species have preferences centered at
\[
\begin{aligned}
(z_1,z_2,z_3)&=((1.8,2.2),(5.5,2.0),(9.5,2.2)),\\
(z_4,z_5,z_6)&=((1.8,8.9),(5.5,8.9),(9.5,8.9)).
\end{aligned}
\]
Their conditional probabilities are
\[
p(s\mid i)=\frac{\exp(-\|x_i-z_s\|^2/[2(2.35)^2])}
{\sum_{s'}\exp(-\|x_i-z_{s'}\|^2/[2(2.35)^2])}.
\]
We simulate 600 independent site visits from $r_i=e_i/\sum_j e_j$
and one categorical species label on each visit, then calculate
order-two Hill numbers from the resulting shares.
For each cap $c\in\{1,2,4,8,16,32,64,128,256,\infty\}$,
weights are defined by \eqref{eq:sim_weights}. The transport
cost matrix contains exact Euclidean distances between the 144
cell centers and is minimized by SciPy's HiGHS linear-programming
solver with exact marginal constraints. Figure~\ref{fig:sim_tradeoff} displays the reweighting candidates
obtained as the cap varies.

The Watarase primary summary uses the November 2025 release of
the eBird Basic Dataset sampling-event records \citep{ebird_ebd}.
The taxon-independent filters are complete, positive-duration
Traveling and Stationary events from May--July 2006--2025 within
20\,km of latitude $36.2140416^\circ$\,N and longitude
$139.6731143^\circ$\,E.
The primary extraction uses a local equirectangular grid:
for latitude $\varphi$ and longitude $\ell$ in radians, put
\[
x=6371.0088\cos(\varphi_0)(\ell-\ell_0),\quad
 y=6371.0088(\varphi-\varphi_0),
\]
where $(\varphi_0,\ell_0)$ is the center given above and cell indices at width $h$\,km are
$(\lfloor x/h\rfloor,\lfloor y/h\rfloor)$.
The independently archived 587-checklist subset, used only for
sensitivity checks, is keyed by sampling-event identifier and has
latitude, longitude, date and group identifier columns. Nonmissing
group identifiers are collapsed to one geographic event; submissions
without a group identifier retain their unique event IDs. Within
each of the 55 nonempty groups, dates and coordinates agree.
For the archived data, the group-collapse and grid comparisons
use $h\in\{0.5,1,2\}$\,km, and a separate row excludes 2025.
The script \texttt{run\_\allowbreak watarase\_\allowbreak sensitivity.py} produces only
aggregate summary rows, never event-level extracts. The archived
sample included six Historical checklists and a fixed operational
taxon-label filter; it is consequently a distinct sample rather
than the 617-event primary set.

Table~\ref{tab:repeat_tradeoff} reports independently regenerated
200-repeat synthetic results with random seed 20261115. Each replicate
draws 600 locations and one categorical label per visit with the same
specified population parameters. The exact $W_1$ is solved by
linear programming on cell-center Euclidean costs. The single
realization, with seed 20261114, is shown in
Table~\ref{tab:sim_tradeoff}. The geometric optimum is generated by
nearest-visited-cell assignment with ties resolved by site index;
within-site weights are spread equally among recorded visits.
\begin{table}[htbp]
\centering\small
\caption{Synthetic reweighting over 200 independent replicates
(seed 20261115), with entries mean (standard deviation). Each
replicate comprises 600 visits.}
\label{tab:repeat_tradeoff}
\begin{tabular}{@{}rrrr@{}}
\toprule
Cap & Effective area (km$^2$) & Species Hill & $W_1$ (km)\\
\midrule
1 & 32.19 (1.56) & 3.21 (0.11) & 3.399 (0.075)\\
8 & 54.59 (2.63) & 3.93 (0.18) & 2.704 (0.130)\\
32 & 63.82 (4.27) & 4.75 (0.28) & 2.016 (0.232)\\
128 & 48.29 (4.74) & 5.50 (0.29) & 1.148 (0.265)\\
Uncapped & 39.52 (4.65) & 5.47 (0.33) & 1.138 (0.248)\\
\bottomrule
\end{tabular}
\end{table}

\section*{Data and code availability}
Reproducible code and synthetic inputs are supplied for the
periodic-population and simulated reweighting experiments, including
the exact transport computations and aggregate-only output files.
The Watarase sampling-event analyses use the November 2025 eBird Basic
Dataset, to be obtained directly from the Cornell Lab of Ornithology
under its terms of access. The supplied sensitivity script accepts
an independently obtained checklist-metadata CSV and emits an
aggregate table. Neither the publication archive nor the code outputs
contain event-level eBird records, geographic point extracts, or
observer identifiers. The 617-event summary and the 587-event sensitivity sample have
distinct extraction criteria and are reported separately.

\section*{Declaration of generative AI and AI-assisted technologies}
During manuscript preparation and development of computational
reproducibility materials, the authors used OpenAI ChatGPT for
assistance with proofreading and code checking.
The authors reviewed the resulting text and calculations and take
full responsibility for the submitted material. The numerical
methods and their settings are described in Section~\ref{sec:watarase}
and Appendix~\ref{app:sim_reproduce}.

\bibliographystyle{plainnat}
\begingroup
\small
\setlength{\bibsep}{0pt}
\bibliography{refs1}

\begin{thebibliography}{54}
\providecommand{\natexlab}[1]{#1}
\providecommand{\url}[1]{\texttt{#1}}
\expandafter\ifx\csname urlstyle\endcsname\relax
  \providecommand{\doi}[1]{doi: #1}\else
  \providecommand{\doi}{doi: \begingroup \urlstyle{rm}\Url}\fi

\bibitem[Amari(2016)]{amari2016information}
Shun-ichi Amari.
\newblock \emph{Information Geometry and Its Applications}, volume 194 of
  \emph{Applied Mathematical Sciences}.
\newblock Springer, Tokyo, 2016.

\bibitem[Amari and Nagaoka(2000)]{amari2000methods}
Shun-ichi Amari and Hiroshi Nagaoka.
\newblock \emph{Methods of Information Geometry}, volume 191 of
  \emph{Translations of Mathematical Monographs}.
\newblock American Mathematical Society and Oxford University Press,
  Providence, RI, 2000.

\bibitem[Angulo et~al.(2021)Angulo, Esquivel, Madrid, and
  Alonso]{angulo2021spatial}
Jos{\'e}~M. Angulo, Francisco~J. Esquivel, Ana~E. Madrid, and Francisco~J.
  Alonso.
\newblock Information and complexity analysis of spatial data.
\newblock \emph{Spatial Statistics}, 42:\penalty0 100462, 2021.
\newblock \doi{10.1016/j.spasta.2020.100462}.

\bibitem[Arditi et~al.(2015)Arditi, Lobry, and Sari]{arditi2015dispersal}
Roger Arditi, Claude Lobry, and Tewfik Sari.
\newblock Is dispersal always beneficial to carrying capacity? {N}ew insights
  from the multi-patch logistic equation.
\newblock \emph{Theoretical Population Biology}, 106:\penalty0 45--59, 2015.
\newblock \doi{10.1016/j.tpb.2015.10.001}.

\bibitem[Benamou and Brenier(2000)]{benamou2000computational}
Jean-David Benamou and Yann Brenier.
\newblock A computational fluid mechanics solution to the {Monge--Kantorovich}
  mass transfer problem.
\newblock \emph{Numerische Mathematik}, 84:\penalty0 375--393, 2000.
\newblock \doi{10.1007/s002110050002}.

\bibitem[Campbell(1966)]{campbell1966extent}
L.~L. Campbell.
\newblock Exponential entropy as a measure of extent of a distribution.
\newblock \emph{Zeitschrift f{\"u}r Wahrscheinlichkeitstheorie und Verwandte
  Gebiete}, 5\penalty0 (3):\penalty0 217--225, 1966.
\newblock \doi{10.1007/BF00533058}.

\bibitem[Cantrell et~al.(2008)Cantrell, Cosner, and Lou]{cantrell2008}
Robert~Stephen Cantrell, Chris Cosner, and Yuan Lou.
\newblock Approximating the ideal free distribution via
  reaction--diffusion--advection equations.
\newblock \emph{Journal of Differential Equations}, 245\penalty0 (12):\penalty0
  3687--3703, 2008.
\newblock \doi{10.1016/j.jde.2008.07.024}.

\bibitem[Cantrell et~al.(2010)Cantrell, Cosner, and Lou]{cantrell2010evolution}
Robert~Stephen Cantrell, Chris Cosner, and Yuan Lou.
\newblock Evolution of dispersal and the ideal free distribution.
\newblock \emph{Mathematical Biosciences and Engineering}, 7\penalty0
  (1):\penalty0 17--36, 2010.

\bibitem[Chao et~al.(2014)Chao, Chiu, and Jost]{chao2014unifying}
Anne Chao, Chun-Huo Chiu, and Lou Jost.
\newblock Unifying species diversity, phylogenetic diversity, functional
  diversity, and related similarity and differentiation measures through {Hill}
  numbers.
\newblock \emph{Annual Review of Ecology, Evolution, and Systematics},
  45:\penalty0 297--324, 2014.
\newblock \doi{10.1146/annurev-ecolsys-120213-091540}.

\bibitem[Cosner(2005)]{cosner2005}
Chris Cosner.
\newblock A dynamic model for the ideal-free distribution as a partial
  differential equation.
\newblock \emph{Theoretical Population Biology}, 67\penalty0 (2):\penalty0
  101--108, 2005.
\newblock \doi{10.1016/j.tpb.2004.09.002}.

\bibitem[Costa(1985)]{costa1985new}
Max H.~M. Costa.
\newblock A new entropy power inequality.
\newblock \emph{IEEE Transactions on Information Theory}, 31\penalty0
  (6):\penalty0 751--760, 1985.

\bibitem[DeAngelis et~al.(2016)DeAngelis, Ni, and
  Zhang]{deangelis2016dispersal}
Donald~L. DeAngelis, Wei-Ming Ni, and Bo~Zhang.
\newblock Dispersal and spatial heterogeneity: single species.
\newblock \emph{Journal of Mathematical Biology}, 72\penalty0 (1):\penalty0
  239--254, 2016.
\newblock \doi{10.1007/s00285-015-0879-y}.

\bibitem[{eBird}(2025)]{ebird_ebd}
{eBird}.
\newblock {eBird Basic Dataset}. version: {EBD\_relNov-2025}.
\newblock Cornell Lab of Ornithology, Ithaca, New York, 2025.
\newblock November 2025 release. Recommended citation supplied with the
  download.

\bibitem[Eguchi(2025)]{eguchi2025maxdiv}
Shinto Eguchi.
\newblock Information geometry for maximum diversity distributions.
\newblock \emph{Sankhya A}, 2025.
\newblock \doi{10.1007/s13171-025-00393-x}.

\bibitem[Falc{\'o} et~al.(2023)Falc{\'o}, Cohen, Carrillo, and
  Baker]{falco2023tissue}
Carles Falc{\'o}, Daniel~J. Cohen, Jos{\'e}~A. Carrillo, and Ruth~E. Baker.
\newblock Quantifying tissue growth, shape and collision via continuum models
  and {Bayesian} inference.
\newblock \emph{Journal of the Royal Society Interface}, 20:\penalty0 20230184,
  2023.

\bibitem[Fithian and Hastie(2013)]{FithianHastie2013}
William Fithian and Trevor Hastie.
\newblock Finite-sample equivalence in statistical models for presence-only
  data.
\newblock \emph{The Annals of Applied Statistics}, 7\penalty0 (4):\penalty0
  1917--1939, 2013.

\bibitem[Freedman and Waltman(1977)]{freedman1977mathematical}
H.~I. Freedman and Paul Waltman.
\newblock Mathematical models of population interactions with dispersal. {I}:
  Stability of two habitats with and without a predator.
\newblock \emph{SIAM Journal on Applied Mathematics}, 32\penalty0 (3):\penalty0
  631--648, 1977.
\newblock \doi{10.1137/0132052}.

\bibitem[Fretwell and Lucas(1969)]{fretwell1969territorial}
Stephen~D. Fretwell and Henry~L. Lucas.
\newblock On territorial behavior and other factors influencing habitat
  distribution in birds.
\newblock \emph{Acta Biotheoretica}, 19\penalty0 (1):\penalty0 16--36, 1969.

\bibitem[Gadgil(1971)]{gadgil1971dispersal}
Madhav Gadgil.
\newblock Dispersal: population consequences and evolution.
\newblock \emph{Ecology}, 52\penalty0 (2):\penalty0 253--261, 1971.
\newblock \doi{10.2307/1934583}.

\bibitem[Gadgil et~al.(2000)Gadgil, Seshagiri~Rao, Utkarsh, Pramod, and
  Chhatre]{gadgil2000participatory}
Madhav Gadgil, P.~R. Seshagiri~Rao, G.~Utkarsh, P.~Pramod, and Ashwini Chhatre.
\newblock New meanings for old knowledge: the {People's Biodiversity Registers}
  program.
\newblock \emph{Ecological Applications}, 10\penalty0 (5):\penalty0 1307--1317,
  2000.
\newblock \doi{10.1890/1051-0761(2000)010[1307:NMFOKT]2.0.CO;2}.

\bibitem[Gilpin and Ayala(1973)]{gilpin1973global}
Michael~E. Gilpin and Francisco~J. Ayala.
\newblock Global models of growth and competition.
\newblock \emph{Proceedings of the National Academy of Sciences of the USA},
  70\penalty0 (12):\penalty0 3590--3593, 1973.

\bibitem[Guo et~al.(2020)Guo, He, and Ni]{guo2020capacity}
Qian Guo, Xiaoqing He, and Wei-Ming Ni.
\newblock On the effects of carrying capacity and intrinsic growth rate on
  single and multiple species in spatially heterogeneous environments.
\newblock \emph{Journal of Mathematical Biology}, 81\penalty0 (2):\penalty0
  403--433, 2020.
\newblock \doi{10.1007/s00285-020-01507-9}.

\bibitem[Gurney and Nisbet(1975)]{gurney1975regulation}
William S.~C. Gurney and Roger~M. Nisbet.
\newblock The regulation of inhomogeneous populations.
\newblock \emph{Journal of Theoretical Biology}, 52\penalty0 (2):\penalty0
  441--457, 1975.
\newblock \doi{10.1016/0022-5193(75)90011-9}.

\bibitem[Hastings(1983)]{hastings1983can}
Alan Hastings.
\newblock Can spatial variation alone lead to selection for dispersal?
\newblock \emph{Theoretical Population Biology}, 24:\penalty0 244--251, 1983.

\bibitem[Hill(1973)]{hill1973diversity}
Mark~O. Hill.
\newblock Diversity and evenness: a unifying notation and its consequences.
\newblock \emph{Ecology}, 54\penalty0 (2):\penalty0 427--432, 1973.
\newblock \doi{10.2307/1934352}.

\bibitem[Holt(1985)]{holt1985population}
Robert~D. Holt.
\newblock Population dynamics in two-patch environments: some anomalous
  consequences of an optimal habitat distribution.
\newblock \emph{Theoretical Population Biology}, 28:\penalty0 181--208, 1985.

\bibitem[Johnston et~al.(2020)Johnston, Moran, Musgrove, Fink, and
  Baillie]{JohnstonEtAl2020}
Alison Johnston, Nick Moran, Andy Musgrove, Daniel Fink, and Stephen~R.
  Baillie.
\newblock Estimating species distributions from spatially biased citizen
  science data.
\newblock \emph{Ecological Modelling}, 422:\penalty0 108927, 2020.
\newblock \doi{10.1016/j.ecolmodel.2019.108927}.

\bibitem[Jordan et~al.(1998)Jordan, Kinderlehrer, and
  Otto]{jordan1998variational}
Richard Jordan, David Kinderlehrer, and Felix Otto.
\newblock The variational formulation of the {Fokker--Planck} equation.
\newblock \emph{SIAM Journal on Mathematical Analysis}, 29\penalty0
  (1):\penalty0 1--17, 1998.

\bibitem[Jost(2006)]{jost2006entropy}
Lou Jost.
\newblock Entropy and diversity.
\newblock \emph{Oikos}, 113\penalty0 (2):\penalty0 363--375, 2006.

\bibitem[Kelling et~al.(2019)Kelling, Johnston, Bonn, Fink, Ruiz-Gutierrez,
  Bonney, Fernandez, Hochachka, Julliard, Kraemer, and
  Guralnick]{KellingEtAl2019}
Steve Kelling, Alison Johnston, Aletta Bonn, Daniel Fink, Viviana
  Ruiz-Gutierrez, Rick Bonney, Miguel Fernandez, Wesley~M. Hochachka, Romain
  Julliard, Roland Kraemer, and Robert Guralnick.
\newblock Using semistructured surveys to improve citizen science data for
  monitoring biodiversity.
\newblock \emph{BioScience}, 69\penalty0 (3):\penalty0 170--179, 2019.
\newblock \doi{10.1093/biosci/biz010}.

\bibitem[Kish(1965)]{kish1965survey}
Leslie Kish.
\newblock \emph{Survey Sampling}.
\newblock John Wiley and Sons, New York, 1965.

\bibitem[Komori et~al.(2020)Komori, Eguchi, Saigusa, Kusumoto, and
  Kubota]{komori2020quasiPPP}
Osamu Komori, Shinto Eguchi, Yusuke Saigusa, Buntarou Kusumoto, and Yasuhiro
  Kubota.
\newblock Sampling bias correction in species distribution models by
  quasi-linear {Poisson} point process.
\newblock \emph{Ecological Informatics}, 55:\penalty0 101015, 2020.
\newblock \doi{10.1016/j.ecoinf.2019.101015}.

\bibitem[Komori et~al.(2023)Komori, Saigusa, and Eguchi]{komori2023sdm}
Osamu Komori, Yusuke Saigusa, and Shinto Eguchi.
\newblock Statistical learning for species distribution models in ecological
  studies.
\newblock \emph{Japanese Journal of Statistics and Data Science}, 6:\penalty0
  803--826, 2023.
\newblock \doi{10.1007/s42081-023-00206-1}.

\bibitem[Kondratyev and Vorotnikov(2020)]{kondratyev2020reaction}
Stanislav Kondratyev and Dmitry Vorotnikov.
\newblock Nonlinear {Fokker--Planck} equations with reaction as gradient flows
  of the free energy.
\newblock \emph{Journal of Functional Analysis}, 278\penalty0 (2):\penalty0
  108310, 2020.
\newblock \doi{10.1016/j.jfa.2019.108310}.

\bibitem[Kondratyev et~al.(2016)Kondratyev, Monsaingeon, and
  Vorotnikov]{kondratyev2016fitness}
Stanislav Kondratyev, L{\'e}onard Monsaingeon, and Dmitry Vorotnikov.
\newblock A fitness-driven cross-diffusion system from population dynamics as a
  gradient flow.
\newblock \emph{Journal of Differential Equations}, 261\penalty0 (5):\penalty0
  2784--2808, 2016.
\newblock \doi{10.1016/j.jde.2016.05.012}.

\bibitem[Lloyd(1967)]{lloyd1967crowding}
Monte Lloyd.
\newblock Mean crowding.
\newblock \emph{Journal of Animal Ecology}, 36\penalty0 (1):\penalty0 1--30,
  1967.
\newblock \doi{10.2307/3012}.

\bibitem[Lou(2006)]{lou2006migration}
Yuan Lou.
\newblock On the effects of migration and spatial heterogeneity on single and
  multiple species.
\newblock \emph{Journal of Differential Equations}, 223\penalty0 (2):\penalty0
  400--426, 2006.

\bibitem[{Ministry of Land, Infrastructure, Transport and
  Tourism}({n.d.})]{mlitWataraseHistory}
{Ministry of Land, Infrastructure, Transport and Tourism}.
\newblock Watarase-yusuichi: overview and history.
\newblock \url{https://www.ktr.mlit.go.jp/tonejo/en/tonejo00081e.html}, {n.d.}
\newblock Accessed 2026-10-25.

\bibitem[Morisita(1959)]{morisita1959dispersion}
Masaaki Morisita.
\newblock Measuring of the dispersion of individuals and analysis of the
  distributional patterns.
\newblock \emph{Memoirs of the Faculty of Science, Kyushu University, Series E
  (Biology)}, 2:\penalty0 215--235, 1959.

\bibitem[Otto(2001)]{otto2001geometry}
Felix Otto.
\newblock The geometry of dissipative evolution equations: the porous medium
  equation.
\newblock \emph{Communications in Partial Differential Equations}, 26\penalty0
  (1--2):\penalty0 101--174, 2001.

\bibitem[Perthame et~al.(2014)Perthame, Quir{\'o}s, and
  V{\'a}zquez]{perthame2014hele}
Beno{\^i}t Perthame, Fernando Quir{\'o}s, and Juan~Luis V{\'a}zquez.
\newblock The {Hele--Shaw} asymptotics for mechanical models of tumor growth.
\newblock \emph{Archive for Rational Mechanics and Analysis}, 212:\penalty0
  93--127, 2014.

\bibitem[Reeve et~al.(2016)Reeve, Leinster, Cobbold, Thompson, Brummitt,
  Mitchell, and Matthews]{reeve2016partition}
Richard Reeve, Tom Leinster, Christina~A. Cobbold, Jill Thompson, Neil
  Brummitt, Sonia~N. Mitchell, and Louise Matthews.
\newblock How to partition diversity.
\newblock \emph{arXiv preprint arXiv:1404.6520}, 2016.
\newblock URL \url{https://arxiv.org/abs/1404.6520}.

\bibitem[R{\'e}nyi(1961)]{renyi1961entropy}
Alfr{\'e}d R{\'e}nyi.
\newblock On measures of entropy and information.
\newblock In \emph{Proceedings of the Fourth Berkeley Symposium on Mathematical
  Statistics and Probability}, volume~1, pages 547--561. University of
  California Press, 1961.

\bibitem[Saigusa et~al.(2024)Saigusa, Eguchi, and Komori]{saigusa2024robustPPP}
Yusuke Saigusa, Shinto Eguchi, and Osamu Komori.
\newblock Robust minimum divergence estimation in a spatial {Poisson} point
  process.
\newblock \emph{Ecological Informatics}, 81:\penalty0 102569, 2024.
\newblock \doi{10.1016/j.ecoinf.2024.102569}.

\bibitem[Savar{\'e} and Toscani(2014)]{savare2014concavity}
Giuseppe Savar{\'e} and Giuseppe Toscani.
\newblock The concavity of {R\'enyi} entropy power.
\newblock \emph{IEEE Transactions on Information Theory}, 60\penalty0
  (5):\penalty0 2687--2693, 2014.
\newblock \doi{10.1109/TIT.2014.2309341}.

\bibitem[Shigesada(1980)]{shigesada1980dispersing}
Nanako Shigesada.
\newblock Spatial distribution of dispersing animals.
\newblock \emph{Journal of Mathematical Biology}, 9:\penalty0 85--96, 1980.
\newblock \doi{10.1007/BF00276037}.

\bibitem[Shigesada et~al.(1979)Shigesada, Kawasaki, and
  Teramoto]{shigesada1979segregation}
Nanako Shigesada, Kohkichi Kawasaki, and Ei~Teramoto.
\newblock Spatial segregation of interacting species.
\newblock \emph{Journal of Theoretical Biology}, 79\penalty0 (1):\penalty0
  83--99, 1979.
\newblock \doi{10.1016/0022-5193(79)90258-3}.

\bibitem[Sibly et~al.(2005)Sibly, Barker, Denham, Hone, and
  Pagel]{sibly2005regulation}
Richard~M. Sibly, Daniel Barker, Michael~C. Denham, Jim Hone, and Mark Pagel.
\newblock On the regulation of populations of mammals, birds, fish, and
  insects.
\newblock \emph{Science}, 309\penalty0 (5734):\penalty0 607--609, 2005.
\newblock \doi{10.1126/science.1110760}.

\bibitem[Stam(1959)]{stam1959some}
A.~J. Stam.
\newblock Some inequalities satisfied by the quantities of information of
  {Fisher} and {Shannon}.
\newblock \emph{Information and Control}, 2\penalty0 (2):\penalty0 101--112,
  1959.

\bibitem[Sullivan et~al.(2009)Sullivan, Wood, Iliff, Bonney, Fink, and
  Kelling]{sullivan2009ebird}
Brian~L. Sullivan, Christopher~L. Wood, Marshall~J. Iliff, Rick~E. Bonney,
  Daniel Fink, and Steve Kelling.
\newblock {eBird}: a citizen-based bird observation network in the biological
  sciences.
\newblock \emph{Biological Conservation}, 142:\penalty0 2282--2292, 2009.

\bibitem[{Tochigi City}({n.d.})]{tochigiWataraseHistory}
{Tochigi City}.
\newblock History of {Watarase-yusuichi} and {Lake Yanaka}.
\newblock
  \url{https://www.city.tochigi.lg.jp/site/watarase-heartland/2877.html},
  {n.d.}
\newblock Accessed 2026-10-25.

\bibitem[V{\'a}zquez(2007)]{vazquez2007porous}
Juan~Luis V{\'a}zquez.
\newblock \emph{The Porous Medium Equation: Mathematical Theory}.
\newblock Oxford University Press, 2007.

\bibitem[Warton and Shepherd(2010)]{WartonShepherd2010}
David~I. Warton and Leah~C. Shepherd.
\newblock Poisson point process models solve the pseudo-absence problem for
  presence-only data in ecology.
\newblock \emph{The Annals of Applied Statistics}, 4\penalty0 (3):\penalty0
  1383--1402, 2010.

\bibitem[Zhang et~al.(2017)Zhang, Kula, Mack, Zhai, Ryce, Ni, DeAngelis, and
  Van~Dyken]{zhang2017carrying}
Bo~Zhang, Alex Kula, Keenan M.~L. Mack, Lu~Zhai, Arrix~L. Ryce, Wei-Ming Ni,
  Donald~L. DeAngelis, and J.~David Van~Dyken.
\newblock Carrying capacity in a heterogeneous environment with habitat
  connectivity.
\newblock \emph{Ecology Letters}, 20\penalty0 (9):\penalty0 1118--1128, 2017.
\newblock \doi{10.1111/ele.12807}.

\end{thebibliography}
\endgroup

\end{document}